\documentclass[a4paper,11pt]{article}
\pdfoutput=1 

\usepackage{jcappub} 

\usepackage[T1]{fontenc} 

\newcommand{\df}{F_K}
\newcommand{\nn}{\nonumber}
\newcommand{\ph}{\phantom}
\usepackage{graphicx}
\usepackage{amssymb}
\usepackage{color}
\usepackage{tensor}
\def\physrep{phys. rep.}
\def\apj{ApJ}
\def\prl{PRL}
\def\mnras{MNRAS}
\def\jcap{JCAP}

\title{\boldmath Consistent cosmological structure formation on all scales in relativistic extensions of MOND}

\author[a,1]{D. B. Thomas\note{Corresponding author.},}
\author[b]{A. Mozaffari}
\author[c]{and T. Zlosnik}

\affiliation[a]{Department of Physics and Astronomy, Queen Mary University of London, UK}
\affiliation[b]{University College London, 
Gower Street, London, WC1E 6BT}
\affiliation[c]{University Of Gda\'{n}sk, Jana Ba{\.z}y\'{n}skiego 8, 80-309 Gda\'{n}sk, Poland}

\emailAdd{dan.b.thomas1@gmail.com}
\emailAdd{a.mozaffari@ucl.ac.uk}
\emailAdd{thomas.zlosnik@ug.edu.pl}

\abstract{
General relativity manifests very similar equations in different regimes, notably in large scale cosmological perturbation theory, non-linear cosmological structure formation, and in weak field galactic dynamics. The same is not necessarily true in alternative gravity theories, in particular those that possess MONDian behaviour (``relativistic extensions'' of MOND). In these theories different regimes are typically studied quite separately, sometimes even with the freedom in the theories chosen differently in different regimes. If we wish to properly and fully test complete cosmologies containing MOND against the $\Lambda$CDM paradigm then we need to understand cosmological structure formation on all scales, and do so in a coherent and consistent manner. We propose a method for doing so and apply it to generalised Einstein-Aether theories as a case study. We derive the equations that govern cosmological structure formation on all scales in these theories and show that the same free function (which may contain both Newtonian and MONDian branches) appears in the cosmological background, linear perturbations, and non-linear cosmological structure formation. We show that MONDian behaviour on galactic scales does not necessarily result in MONDian behaviour on cosmological scales, and for MONDian behaviour to arise cosmologically, there will be no modification to the Friedmann equations governing the evolution of the homogeneous cosmological background. We comment on how existing N-body simulations relate to complete and consistent generalised Einstein-Aether cosmologies. The equations derived in this work allow consistent cosmological N-body simulations to be run in these theories whether or not MONDian behaviour manifests on cosmological scales.}

\begin{document}
\maketitle
\flushbottom

\section{Introduction}
The current cosmological paradigm, $\Lambda$CDM is broadly successful \cite{planck}, although it is has some ongoing small scale problems and tensions (see e.g. \cite{1606.07790,1707.04256,2103.01183,2105.05208}). This paradigm contains two hypothesised forms of matter: cold dark matter and dark energy. The existing observational evidence for these is solely gravitational in nature, so it is natural to consider whether modified gravitational laws (rather than Einstein's General Relativity (GR)) could instead be responsible for these observations, see e.g. \cite{1106.2476} for a comprehensive review. A long-running suggestion to account for some of the phenomenology attributed to dark matter is MOdified Newtonian Dynamics (MOND) \cite{milgrom1,milgrom2,milgrom3}. See \cite{milgromreview,famaeyreview,1206.6231,banikreview} for reviews of the theory and observational successes and challenges of MOND.

The successes and failures of MOND (and particularly how it fares versus the cold dark matter of  $\Lambda$CDM) are an ongoing debate in the literature (see e.g. \cite{1510.01369,1609.05917,1811.05260,1901.05966,2009.11525,2109.04745,2112.00026,2201.11752,2202.01221,2210.13961} for a non-exhaustive list of recent topics and discussions). In this work we do not take a position advocating for or against MOND. Rather, we take the view that until there is non-gravitational evidence for dark matter, it is worth testing and developing the different paradigms in an attempt to falsify each of them. However, to do this we need to understand cosmological structure formation on all scales\footnote{Note that throughout, we use ``cosmological structure formation on all scales'' to mean all scales where a perturbed FLRW metric with weak fields is a reasonable description of the spacetime (a range of scales from super horizon scales to approximately scales of order Mpc), and by ``non-linear scales'' we mean scales of around 10Mpc and below, where the cosmological density contrast $\delta$ becomes greater than 1.} in a MONDian cosmology and there is as yet no complete and consistent MONDian cosmology that makes clear predictions on all cosmological scales and that can be fully and fairly tested against the $\Lambda$CDM paradigm.
Partly this is difficult because MOND itself is not a theory of gravity, it is a phenomenological description of how the gravitational laws might behave in certain limits. To be a complete theory, MONDian behaviour must be embedded in a ``relativistic extension'', several candidates for which have been investigated \cite{bekensteinmilgrom1984,0403694,1201.2759,0912.0790,2007.00082,2109.13287}. See references in \cite{2007.00082} for a more comprehensive list of different theories in which MONDian behaviour can arise in different ways.

An additional difficulty is that studying cosmological structure formation requires looking at several different regimes, none of which are the weak-field galactic dynamics regime for which MOND was originally proposed. MONDian behaviour is often expressed as a modification of the Poisson equation\footnote{Note that in this paper we consider MOND as a modification of gravity not as modified inertia.} in this regime, but here we run into the issue that the Poisson equation is something that is a little convenient in GR+$\Lambda$CDM: a very similar equation arises in conceptually quite different regimes, notably large scale cosmological perturbation theory, non-linear cosmological structure formation, and in weak field galactic (and smaller scale) dynamics. The equations in these regimes are not necessarily so similar in other theories of gravity, and this particularly applies to the case of relativistic extensions of MOND. These theories are usually studied piecemeal in different regimes, with different assumptions and frameworks in each, and sometimes even with the available freedom in the theories being chosen differently in different regimes. This disjointed approach is theoretically problematic as it is unclear if the different choices made in different regimes can be simultaneously realised in the same universe. It is also practically problematic since it means it is unclear how to run cosmological simulations when the different limits are treated so differently, due to the range of scales that is being covered by these simulations. In particular these simulations need to have a cosmological background and large scale perturbations that are consistent with each other, and with the smaller scale non-linear behaviour. Studying cosmological structure formation in GR is more straightforward because of its similarity across the different regimes.

Cosmological N-body simulations up to scales of $512h^{-1}\text{Mpc}$ have previously been run with some form of MONDian Poisson equation \cite{0109016,0303222,0809.2899,1104.5040,1309.6094,1410.3844,1605.03192}, however due to the issues noted above it is somewhat unclear how these simulations relate to specific relativistic extensions of MOND. From the other side, the equations that govern cosmological structure formation on all scales have not been derived for any relativistic extensions of MOND; in particular, equations that can be used to run cosmological N-body simulations for a specific theory. Another use of such equations would be to examine on which scales involved in cosmological structure formation MONDian behaviour can arise, and if MONDian behaviour on these scales necessarily arises from having MONDian behaviour on galactic scales.

In order for the MOND paradigm to develop into sufficient maturity to be fully and consistently compared against $\Lambda$CDM, these issues need to be addressed. In this work we examine these issues using the post-Friedmann approach \cite{postf,thesis}; this is a weak field post-Newtonian-like expansion designed to work in a cosmological setting, so it is ideal for such a study. In addition to GR+$\Lambda$CDM, it has previously been applied to Hu-Sawick \cite{husawicki} $f(R)$ gravity \cite{postffr} and used to create a model-independent approach to modified gravity that applies on all cosmological scales \cite{theorypaper,simpaper}. We create a method for deriving the equations that govern cosmological structure formation (the evolution of inhomogeneities) on all scales that is inspired by the observations in \cite{theorypaper}. In particular, the process used to derive the all-scales equations partly draws on the conditions in GR+$\Lambda$CDM that cause the simulation situation to be relatively simple, such as the absence of a regime where neither perturbation theory or the Newtonian limit apply (as discussed in \cite{theorypaper}). We apply this method to generalised Einstein-Aether (GEA) theories \cite{0607411,Dai:2008sf, 0711.0520, 1002.0849,0709.4581,1711.09893, 1707.06508,1811.07805} as an illustrative example of how a relativistic extension of MOND can be examined more holistically over the full range of cosmological scales. This process gives a single set of equations for running consistent cosmological simulations, including an expansion history, large scale behaviour, and small scale behaviour that are consistent with each other. We examine how MONDian behaviour can arise in the resulting structure formation and if and how this relates to MONDian behaviour on galactic scales, as well as what this means for relating existing cosmological MOND N-body simulations to GEA cosmologies. We also examine how one can check the assumptions that underly our derivation.

This paper is laid out as follows: in section \ref{sec_recap} we elaborate on the theoretical context and briefly recap some details of GEA theories and the post-Friedmann approach. In section \ref{sec_derivation} we apply our approach to GEA theories to construct equations that describe cosmological structure formation on all scales in these theories. We discuss some features of these equations in section \ref{sec_discussion}, notably if and how MONDian behaviour can arise cosmologically. We conclude in section \ref{sec_conc}.


\section{Theoretical context}
\label{sec_recap}
In this section we discuss some of the issues around MOND N-body simulations, relating them to relativistic extensions of MOND, and why we need equations for these theories that apply on all scales. We then briefly recap GEA theories (mostly following \cite{0607411, 0711.0520}), and the post-Friedmann approach (mostly following \cite{postf,theorypaper}), and in the latter we explain the method that will be used to derive the equations for structure formation that apply on all cosmological scales.

\subsection{(MOND) N-body simulations and the need for coherent and complete cosmologies on all scales in relativistic extensions of MOND}
The equations governing cosmological structure formation in $\Lambda$CDM+GR are very similar in the linear perturbation limit and in the non-linear Newtonian limit, and there aren't any scales where the leading order dynamics are not well described by one of these two limits (see e.g. the discussion in \cite{theorypaper} and references therein). As a result, it is relatively straightforward (from a physics, if not a computational perspective) to run N-body simulations that cover a wide range of scales, from super horizon scales all the way down to scales where the density fluctuations are large.

\begin{figure}
\centering
\includegraphics[width =12cm]{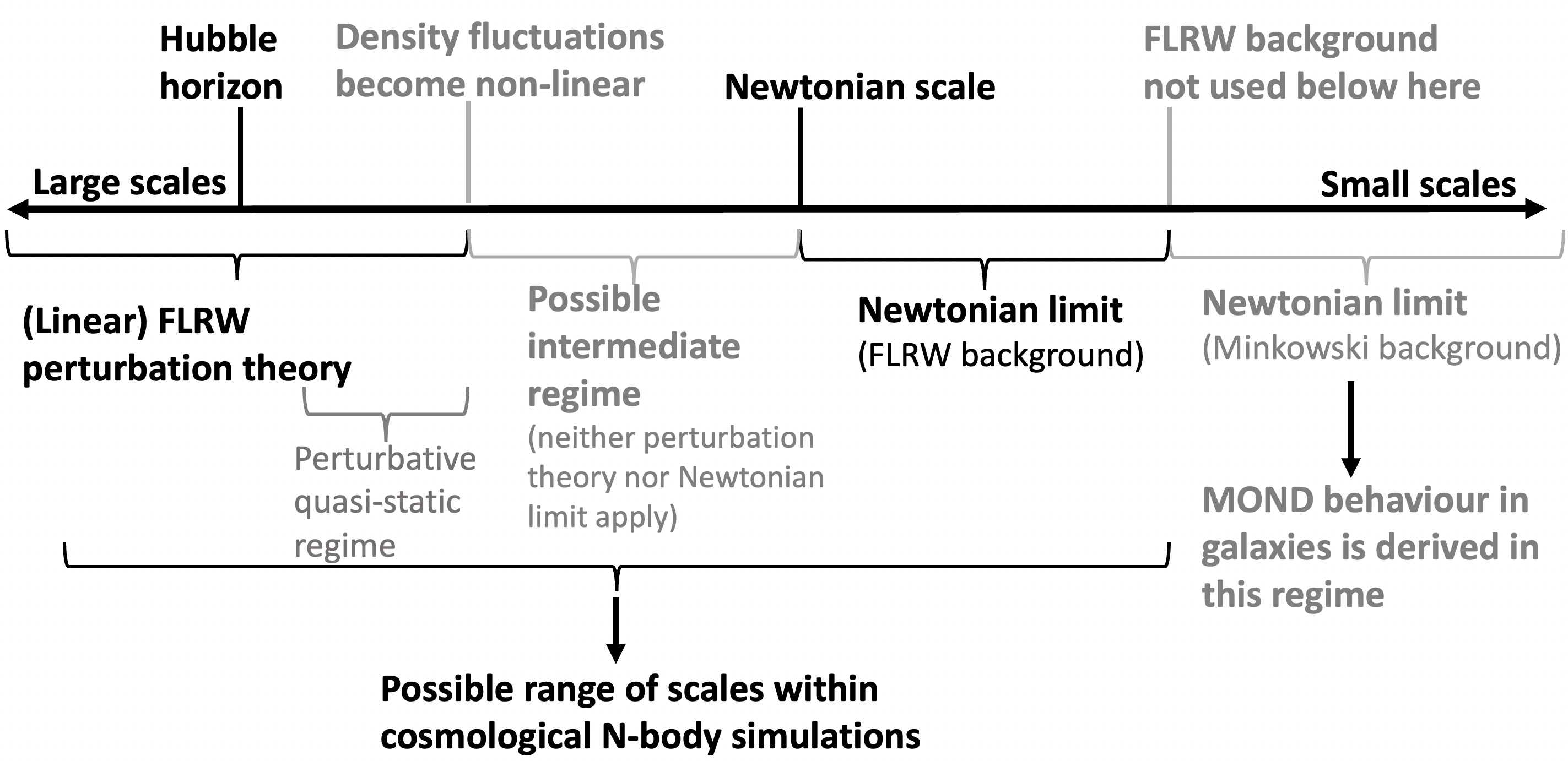}
\caption{Schematic showing the different conceptual regimes referred to throughout the paper, the scales they cover relative to each other, and how this relates to MOND and cosmological simulations. The Newtonian scale denotes the scale below which the Newtonian limit of a relativistic theory is a good approximation to the full equations. Note that there is no intermediate regime in $\Lambda$CDM: in this case the perturbation and Newtonian regimes overlap. Similarly, the perturbative quasi-static regime typically only exists in cosmologies with no intermediate regime.
}
\label{fig_scales}
\end{figure}

The situation is typically more complicated in modified gravity theories, where the different assumptions and properties of the different regimes means that the equations can differ more between the different regimes. This issue is particularly pronounced for relativistic extensions of MOND for two reasons. Firstly because different limits of the theories are typically studied independently, which sometimes even includes choosing the undetermined freedom in the theories differently in different regimes.\footnote{For the Generalised Einstein Aether theories considered later in this work, this freedom is the function $F(K)$ and its form for different ranges of $K$.} The second is that the cosmological regimes have different assumptions to the galactic limit in which MONDian behaviour is shown to arise in these relativistic theories. To the authors' knowledge, a single consistent description covering all of the scales for cosmological structure formation has not previously been derived for a relativistic extension of MOND. Instead, behaviour is typically extrapolated from galactic scales into the non-linear regime of cosmological structure formation (as discussed in the next paragraph); such extrapolations involve moving between regimes in which particular limits of the theory are studied and the freedom is chosen in a particular way. To aid the reader, a schematic illustration of the different scales referred to in this work is show in figure \ref{fig_scales}. This figure shows the different cosmological and non-cosmological regimes, as well as the range of regimes spanned by cosmological N-body simulations. In this work for simplicity we assume that there is a lower limit to the range of scales on which the use of the FLRW metric, and thus the derivation later in this paper, is valid.\footnote{ This means that we always work within a cosmological context and do not consider for example whether one can always move to FLRW co-ordinates even on smaller (e.g. galactic) scales as carried out for example in \cite{ppnc1,ppnc2}). If such transformations are possible in the theories we consider here then this work can be extended to include galactic behaviour and constrain these theories further from the point of view of requiring consistent behaviour on all scales.} As such, the backgrounds of the two Newtonian limits differ, and both the additional time dependence of the FLRW background and that it is not a vacuum solution (but Minkowski is) could create differences for other theories of gravity.

From the simulation side, since MOND is often taken to be a modification to the Newtonian Poisson equation, one common route is to modify N-body simulations to include the MONDian Poisson equation. This is valid on galactic scales where a simple weak field (in a Minkowski background) derivation of MONDian behaviour can be carried out on a particular relativistic extension of MOND. Here, we focus on cosmological-type simulations on larger scales, rather than simulations on smaller scales focused around simulating galaxy formation and evolution. Previous works \cite{0109016,0303222,0809.2899,1104.5040,1309.6094,1410.3844,1605.03192} have been run with box sizes varying from $32 h^{-1}$Mpc to $512h^{-1}$Mpc, all of which involve scales that will be evolving according to linear and non-linear fluctuation behaviour in a cosmological background.

With ``cosmological'' box sizes such as these, it is less clear cut whether MONDian behaviour necessarily arises even in theories that have MONDian behaviour on galactic scales, because cosmological structure formation is a different conceptual regime. Again, just because the GR equations are similar in the different regimes, doesn't mean that this will be the case for other theories. This question has not previously been addressed\footnote{Although we note that careful thought is given in \cite{0109016,0303222,0809.2899,1104.5040,1309.6094,1410.3844,1605.03192} to the fact that the underlying covariant cosmology is not known, requiring assumptions such as the validity of the Friedmann equations for the background expansion, assuming that MONDian effects only apply to the peculiar acceleration, and the validity of the initial conditions that are typically used in $\Lambda$CDM N-body simulations. Since the method in this paper creates a consistent cosmology on all scales and times, these questions will be effectively answered by the same framework, although these are not the primary questions we are investigating.}. This issue is compounded by the aforementioned problems about the different cosmological regimes that are spanned by an N-body simulation. For example, an (effective) modified $G$ in the Poisson equation on linear cosmological scales is a generic prediction of modified gravity theories (see e.g. \cite{1106.2476}), so such an effect may be required on the larger (linear) scales in MONDian cosmological simulations. Such a modification in linear theory needs to be understood in terms of how it relates to the branches of the MOND Poisson equation in non-linear structure formation, whether it arises in one or both of these branches, and therefore how it should arise in a cosmological N-body simulation.

For concreteness, when we refer to galactic MONDian behaviour in this work we are referring to a Poisson equation given by
\begin{align}
\vec{\nabla}_{(p)}\cdot\left(\mu\left(\frac{|\vec{\nabla} \Phi|}{a_0} \right)  \vec{\nabla}_{(p)}\Phi\right)&=4\pi G \rho \label{eqn_mond}\\
\mu(x)&=1 \quad\quad\text{ (Newtonian branch)}\\
\mu(x)&=x\quad \quad \text{ (deep MOND branch),}
\end{align}
where $\rho$ is the matter density and $\Phi$ is the Newtonian potential. Since MOND was proposed in a galactic context, when applying it cosmologically one has to make a choice about whether the $a_0$ function is allowed to vary over cosmic history, and whether the quantities that appear in the MOND Poisson equation (such as $\rho$, $\Phi$ and $\nabla$) are the physical or comoving quantities\footnote{Note that the subscript $_p$ on $\vec{\nabla}$ in equation (\ref{eqn_mond}) denotes that this is a derivative with respect to a physical co-ordinate; $\vec{\nabla}$ throughout the rest of this paper is a derivative with respect to a comoving co-ordinate.}. We make a small generalisation of MONDian behaviour to include an extra function ($\gamma(a)$) of the scale factor $a$ to account for these different possibilities and choices
\begin{eqnarray}
\label{eqn_cosmomond}
\vec{\nabla}\cdot\left(\mu\left(\frac{|\vec{\nabla} \Phi|}{\gamma(a)a_0} \right)  \vec{\nabla}\Phi\right)=4\pi G a^2 \rho \text{.}
\end{eqnarray}
Behaviour matching equation \ref{eqn_cosmomond} will be the behaviour we refer to as cosmological MOND behaviour when examining a specific relativistic extension of MOND later in the paper. We note that this simple form does not quite encompass all of the possible choices, but it covers the main ones and is reasonably general whilst maintaining a fairly simple form. This form is similar to that adopted for cosmological MOND N-body simulations (e.g. \cite{0809.2899,1605.03192}).

To summarise, the following are still open questions: what does a complete picture of cosmological structure formation look like in relativistic extensions of MOND, how do we run consistent cosmological N-body simulations for these theories, how do said simulations relate to existing cosmological MOND simulations, and does MONDian behaviour necessarily arise on cosmological scales if it arises on galactic scales (and if it does appear, then in which scales, times and locations does it appear)? These issues apply to analytic examinations of structure formation in MOND as well (e.g \cite{9710335,0011439}): if a particular relativistic extension of MOND contains MOND on galactic scales but not cosmological scales, then the general lessons drawn from studying cosmological structure formation in MOND will not apply to these theories; conversely cosmological structure formation potentially being found to not be compatible with these studies does not necessarily therefore rule out MOND phenomenology on galactic scales.

To solve these issues in a particular theory, we need to derive equations that describe the evolution of cosmological inhomogeneities on all scales. In this work, we propose a method to do so and illustrate it by applying it to GEA theories as a case study.

\subsection{Generalised Einstein Aether theories}
GEA theories are theories with an additional timelike vector field, $A^\mu$, with an action given by
\begin{equation}
S=\int\sqrt{-g}d^4x\bigg( \frac{c^4}{16\pi G}\left[\left(R-2\Lambda \right)+ M^2F(K)+\lambda \left(A^\mu A_\mu +1 \right)\right]+ \mathcal{L}_m\bigg)
\end{equation}
where $\lambda$ is the Lagrange multiplier ensuring that the Aether field $A^\mu$ has a timelike unit norm and the function $F(K)$ is free and can be chosen to get different phenomenology. The scalar $K$ is given by
\begin{eqnarray}
K=M^{-2}K\indices{^\mu^\nu_\alpha_\beta} \nabla_\mu A^\alpha\nabla_\nu A^\beta\nonumber\\
K\indices{^\mu^\nu_\alpha_\beta}=c_1g^{\mu \nu}g_{\alpha\beta}+c_2\delta^\mu_\alpha \delta^\nu_\beta+c_3\delta^\mu_\beta\delta^\nu_\alpha\text{,}
\end{eqnarray}
The $c_i$ are constants, and are constrained to be $c_1<0$, $c_2\leq0$ and $c_1+c_2+c_3\leq0$. For later convenience we also define $\alpha=c_1+3c_2+c_3$. There is also a possible `$c_4$' contribution $\mathcal{K}^{\mu \nu}_{\ph{\mu\nu}\alpha\beta}=c_4 A^\mu A^\nu g_{\alpha \beta}$, see \cite{1707.06508} for a discussion. For simplicity, we drop this term and concentrate on theories that have been shown to deliver MOND on galactic scales \cite{0607411, 0711.0520}. The method presented later can accommodate non-zero $c_4$ if required. For these theories, the Einstein equations are given by
\begin{align}
G_{\mu\nu}&=8\pi G T^{(m)}_{\mu\nu}+T^{(GEA)}_{\mu\nu}\nonumber\\
T^{(GEA)}_{\alpha\beta}&=\frac{1}{2}\nabla_\gamma\biggl(F_K \left (J_{(\alpha} {}^\gamma A_{\beta )} -J^\gamma_{(\alpha}A_{\beta )}-J_{(\alpha \beta)}A^\gamma\right) \biggr)-F_K Y_{(\alpha\beta)}+\frac{1}{2}g_{\alpha\beta}M^2F+\lambda A_\alpha A_\beta\nonumber\\
Y_{\alpha \beta}&=c_1\biggl((\nabla_\alpha)(\nabla_\beta A^\nu) -(\nabla_\nu A_\alpha)(\nabla^\nu A_\beta)\biggr)\nonumber\\
J\indices{^\mu_\nu}&=\left(K\indices{^\mu^\alpha_\nu _\beta} +K\indices{^\alpha ^\mu_\beta _\nu}\right)\nabla_\alpha A^\beta\nonumber \text{,}
\end{align}
with $F_{K}=\frac{dF}{dK}$ and with $T^{(m)}_{\mu\nu}$ being the stress energy tensor of matter. The vector field equation is given by
\begin{eqnarray}
\nabla_\alpha\left(F_{K} J\indices{^\alpha_\beta} \right)&=&2\lambda A_\beta\text{.}
\end{eqnarray}
%
%

GEA theories have been studied extensively in terms of their cosmological background and linear perturbations \cite{0607411, 0709.4581,0711.0520,1002.0849,1711.09893, 1707.06508,1811.07805}, and also in terms of their gravitational wave propagation \cite{1710.06394,1802.09447}. These theories have also been looked at in the weak field limit \cite{0607411, 0711.0520} that describes the Solar System and galaxies. Notably, the free function $F(K)$ is usually chosen differently in these different regimes in order to get MONDian behaviour in galaxies and also get dark-energy-like cosmological backgrounds. This choice is made on the assumption that there is a single (complicated) $F(K)$ that has different behaviour for different ranges of $K$, and the ranges of values that $K$ takes in the different regimes are disjoint. However, the validity of this assumption is less clear if we include (non-linear) cosmological structure formation in which MONDian behaviour may arise on scales that are described by a FLRW background. This is one of the issues that our approach will resolve.

Note that we are not positing that GEA theories themselves are necessarily strong contenders to $\Lambda$CDM, merely using them as a reasonably well-studied example to illustrate how our method can be used to derive complete and consistent cosmologies in (and insight into) relativistic extensions of MOND.

\subsection{Post-Friedmann formalism}
The post-Friedmann formalism is a post-Newtonian-like expansion of the Einstein equations (or modified Einstein equations) in powers of the speed of light $c$, altered compared to a ``Solar-System'' type expansion in order to apply to a FLRW cosmology \cite{postf, thesis}. The starting point of the post-Friedmann approach is the perturbed FLRW metric in Poisson gauge\footnote{The Poisson gauge is one of the few cosmological gauges that is valid on scales where the density contrast isn't small \cite{cliftongauge}.}, which is expanded up to order $c^{-5}$,
\begin{eqnarray}
 g_{00}&=&-\biggl(1-\frac{2U_N}{c^2}+\frac{1}{c^4}\left(2U^2_N-4U_P \right)\biggr)\\
g_{0i}&=&-\frac{aB^N_i}{c^3}-\frac{aB^P_i}{c^5}\\
g_{ij}&=&a^2\biggl(\left(1+\frac{2V_N}{c^2}+\frac{1}{c^4}\left(2V^2_N+4V_P \right) \right)\delta_{ij} +\frac{h_{ij}}{c^4}\biggr) \rm{.}
\end{eqnarray}
The two scalar potentials have each been split into their leading order (Newtonian) ($U_N$,$V_N$) and higher order ($U_P$,$V_P$) components. The gauge freedom is chosen such that the vector potential appears in the $0i$ part of the metric, and this has also been split up into $B^N_i$ and $B^P_i$. The three-vectors $B^N_i$ and $B^P_i$ are both divergence-less, $B^N_{i,i}=0$ and $B^P_{i,i}=0$. In addition, the tensor perturbation $h_{ij}$ is transverse and trace-free, $h^i_i=h^{,i}_{ij}=0$. Time derivatives are associated with a factor of $\frac{1}{c}$. The matter content (in addition to a possible cosmological constant) is taken to be pressure-less dust\footnote{This is a sufficient description of baryons on most cosmological scales, and is not broken sufficiently on small scales to jeopardise the expansion in $c$.}, the four-velocity of which is used to construct the energy-momentum tensor, which is also expanded in powers of $c$. The parameters describing the pressure-less dust fluid are the background density $\bar{\rho}$, the density contrast $\delta$, and the peculiar velocity $v_i$. Crucially, this setup doesn't require the density contrast to be small, unlike standard (linear) cosmological perturbation theory, which allows the approach to be used on small scales. To obtain the equations that apply on all cosmological scales we will change variables to the ``re-summed potentials'' \cite{postf,theorypaper}
\begin{eqnarray}
\psi_P&=&-V_N-\frac{2}{c^2}V_P\\
\phi_P&=&-U_N-\frac{2}{c^2}U_P\\
\vec{\omega}&=&\vec{B}_N+\frac{2}{c^2}\vec{B}_P\text{.}
\end{eqnarray}
For full details of the post-Friedmann formalism, see \cite{postf}.

\subsubsection{Cosmological structure formation on all scales}
The post-Friedmann formalism was used in \cite{postf} to construct equations that apply on \textit{all} cosmological scales, including both the large and small scale limits and a possible ``intermediate'' regime where neither linear perturbation theory nor the Newtonian limit\footnote{Throughout this work we will use ``Newtonian limit'' to denote the equation of motions in a theory of gravity that occur at leading order in this $\frac{1}{c}$ expansion, and we reserve ``Newtonian branch'' and ``MONDian branch'' for discussions of which limit of the MOND Poisson equation is active in a particular context. I.e. ``Newtonian limit'' does not necessarily mean that the equations of motion are those of Newtonian gravity.} is sufficient for calculating cosmological structure formation. These equations are quite complicated, however they can be simplified \cite{theorypaper} under the assumption that there is no intermediate regime, i.e. that all scales of interest are described by either linear perturbation theory or the cosmological Newtonian limit. This assumption holds for $\Lambda$CDM and at least some modified gravity theories \cite{theorypaper,postffr}. The simplified equations describe the evolution of cosmological homogeneities on every scale from super-horizon scales to the regime of non-linear cosmological structure formation as long as the assumption that there is no intermediate regime holds. For a $\Lambda$CDM+GR cosmology, these simple all scales equations are given by equations $(3.20a)-(3.20d)$ in section IV of \cite{theorypaper} .

In this work we build on the idea of \cite{theorypaper} to present a method for deriving equations that apply on all scales in a specific theory of modified gravity. In \cite{theorypaper} the simplified all scales equations are arrived at by expanding to $c^{-5}$ order, converting to the re-summed potentials, and then removing certain terms from the equations: specifically the terms that are \textit{both} ``structurally non-linear'' and beyond leading order in the $\frac{1}{c}$ expansion. These same equations can be arrived at with a simpler method, namely for each equation derive the two limiting cases, write them in terms of the re-summed potentials, and then combine the two equations into a single equation that contains both limits (and no more). This is the method we will follow here, and it results in a single consistent set of equations that describe the background, and fluctuations on all cosmological scales in these theories. By definition this is equivalent to constructing the full set of master equations up to order $c^{-5}$, converting to the re-summed potentials, and removing the terms that associated only with an intermediate regime where neither the linear perturbation theory or Newtonian limit applies, however it is a much easier computational process. This process amounts to ``stitching together'' the two limits, which works because they are both weak field expansions and thus can both be written in terms of the re-summed potentials.

This method is not universally valid for any cosmology or matter content. By construction, this derivation only applies in a matter or $\Lambda$ dominated cosmology, i.e. the dominant clustering component in the universe needs to be pressureless dust. Since the method followed here yields the same equations as the method laid out in \cite{theorypaper}, the remaining requirements for this method to be valid are largely given by the set of criteria in section V.A of \cite{theorypaper}. These requirements can be approximately summarised as follows: only scalar metric potentials are important for cosmological structure formation, the two limits (linear perturbation and Newtonian) are valid, and every relevant scale is well-described by one (or both) of the two limits (in terms of figure \ref{fig_scales} this means that there is no ``intermediate regime'' and thus that the Newtonian and perturbative limits overlap). The first of these is not required in the GEA context. We will examine the remaining requirements for the GEA case in section \ref{sec_conditions}, here we just note that in any cosmological N-body simulation, the vector potential $\vec{\omega}$ can be used as a quantitative test of the Newtonian limit assumption \cite{Bruni:2013mua,longerpaper,postffr,theorypaper}. As such, this test is one of the tests for this method, and the required equation to perform this test is naturally derived as part of the method; in the GEA case this is equation (\ref{eqn_vector}).

\section{Derivation of all-scales equations in GEA theories}
\label{sec_derivation}
As described in the previous section, this derivation has two parts: finding the leading order post-Friedmann equations, and then combining them with the linear perturbation equations.

\subsection{Applying the post-Friedmann approach to GEA theories}
To begin, we need to decide how to expand the vector field $A^\mu$ in the post-Friedmann $\frac{1}{c}$ expansion. We do this as follows
\begin{align}
A^0=1+\frac{U_N}{c^2}, \quad A^i=\frac{\beta^i}{c^3}\nonumber\\
\end{align}
and for $A_{\mu}$ it follows that:

\begin{align}
A_0=-1+\frac{U_N}{c^2},\quad
A_i=a^2\frac{\beta_i}{c^3}-\frac{aB_i}{c^3}\text{,}
\end{align}
where $\beta_i=\delta_{ij}\beta^j$. Hence the homogeneous background part of the aether field is given by $\bar{A}^\mu=(1,0,0,0)$ as usual. Our choice of expansion for the perturbations is justified by looking at the Lagrange multiplier equation (ensuring $A^\mu$ is timelike) order by order. It is fairly straightforward to expand the perturbation to $A^0$: it typically behaves like a scalar field, which are usually expanded in even powers of $c$ and, from the Lagrange multiplier equation, in the weak field limit it is expected to be equal to $U_N$. Both of these considerations are satisfied by the leading order perturbation to $A^0$ being of order $c^{-2}$. Assuming that the spatial perturbation to the vector field is not of lower order in $c$ than the time perturbation, then the Lagrange multiplier equation at $c^{-3}$ order sets any possible $c^{-3}$ contribution to the time perturbation to be zero. Then the only question is the spatial perturbation: the Lagrange multiplier at $c^{-5}$ order requires
\begin{equation}
\beta^{i(2)}\left(aB^N_i+2a^2\beta^{(3)}_{i} \right)=0\text{,}
\end{equation}
where $\beta^{i(n)}$ refers to the part of $\beta^i$ that comes with an attached power of $c^n$. If $\beta^{i(2)}\neq0$ then one would see that the leading order terms of the spatial part of the vector equation force this term to either be zero or satisfy a strange constraint\footnote{$c_1\beta^{(2),j}_{i,j}+c_2\beta^{(2)j}_{,ij}+c_3\beta^{(2)j}_{,ij}=0$}, so we choose $\beta^{i(2)}=0$; the equations derived under this choice are not consistent with $aB^N_i+2a^2\beta_{i(3)} =0$. It is not surprising for the pure vector part of the spatial perturbation that there is a $c^{-3}$ factor, as the pure vector metric potential is also $c^{-3}$.

In order to be as general as possible with the freedom allowed in GEA theories we do not specify a form of $F(K)$ here, or assign a power of $c$ to $M$ in the derivations. Instead we initially assume that $K$, $F(K)$ and $F_K$ can be sensibly expanded, and therefore in these equations $M^2F(K)$ and $F_K$ terms are taken to represent the leading order part. Terms denoted $F(\bar{K})$ or $\bar{F}_K$ mean only the homogeneous part of any such leading order term. In section \ref{sec_allscales2} we examine the issue of assigning a power of $c$ to $M$ and what this means for the expansion and the resulting equations, as well as how this relates to possible choices of $F(K)$. We then consider specific $F(K)$ that might lead to MONDian behaviour in section \ref{sec_discussion}.

With these choices we can now calculate the leading order equations. We include calculations of some of the intermediate quantities in appendix \ref{app_gory} for the interested reader and to ensure that others can reproduce our results if required. The Lagrange multiplier is calculated from the time part of the vector field equation (throughout, both $\dot{x}$ and $\partial_T x$ denote differentiation with respect to time)
\begin{equation}
\lambda=-3\frac{c_2}{c^2}\dot{F}_K H+\frac{c_3F_{K,i}U_{N,i}}{a^2c^2}+\frac{F_k}{c^2}\bigg(\frac{c_3 U_{N,kk}}{a^2}+3\alpha H^{2}-6c_{2}H^{2}-3\frac{c_2\ddot{a}}{a} \bigg) \text{,}
\end{equation}
and $K$ is given by
\begin{equation}
K=3\frac{\alpha}{M^2c^2}H^{2}+2\frac{\alpha\beta^i_{,i}}{M^2 c^4}H-c_1\frac{U_{N,k}U^{,k}_N}{M^2a^2c^4}+6\frac{\alpha U_{N}}{M^2c^4}H^{2}+\frac{6\alpha\dot{V}_N}{M^2c^4} H\text{.}
\end{equation}
The homogeneous part of $T^{(GEA)}_{\mu\nu}$ (the GEA contributions to the Einstein equations) at leading order (up to $c^{-3}$) is given by
\begin{eqnarray}
&&\bar{T}^{(GEA)}_{0 0}=-\frac{1}{2}M^2F(\bar{K})+3\bar{F}_K\frac{\alpha}{c^2}H^{2}\nonumber\\
&&\bar{T}^{(GEA)}_{i j}= \delta_{ij}\frac{a^{2}}{c^2}\bigg(\bar{F}_K\left(-\frac{\ddot{a}}{a}\alpha-2H^{2}\alpha\right)+\frac{1}{2}a^2M^2F(\bar{K})- \dot{\bar{F}}_Ka\dot{a}\alpha \bigg)\text{,}
\end{eqnarray}
and the inhomogeneous parts are given by
\begin{align}
T^{(GEA)}_{0 0}&=-\frac{1}{2}M^2\left(F_K-\bar{F}_K \right)+3F_K\frac{\alpha}{c^2}H^{2}+c_1F_K\frac{U^{,i}_{N,i}}{a^2 c^2}+c_1F_{K,i}\frac{U_{N,i}}{a^2c^2}+\frac{1}{c^2}U_NM^2F\nonumber\\
T^{(GEA)}_{0 i} &=F_Kc_1\frac{U_{N,i}}{c^3}H+\frac{\dot{F}_{K}c_1U_{N,i}}{c^3}+\frac{F_K c_1\dot{U}_{N,i}}{c^3}-\frac{aB_i}{2c^3}M^2F\nn\\
&+\frac{(c_1-c_3)}{2c^3}\bigg(F_K\bigg(\left(\beta^{,l}_i-\beta^l_{,i} \right)+\frac{1}{a} \left(B^l_{,i}-B^{,l}_i \right)\bigg)\bigg)_{,l}
\nonumber\\
T^{(GEA)}_{i j}&= \delta_{ij}\frac{a^{2}}{c^2}\bigg(\left(F_K-\bar{F}_K \right)\left(-\frac{\ddot{a}}{a}\alpha-2H^2\alpha\right)+\frac{1}{2}M^2\left(F_K-\bar{F}_K \right)- (\dot{F}_K-\dot{\bar{F}}_K) H\alpha\bigg)+\frac{a^2}{c^2}V_N\delta_{ij}M^2F\text{.}
\end{align}
The homogeneous equations are given by
\begin{align}
H^2\left(1-\alpha \bar{F}_K \right)&=\frac{8\pi G \rho}{3}+\frac{\Lambda}{3}-\frac{1}{6}M^2F(\bar{K})\nn\\
-2\frac{\ddot{a}}{a}-H^2&=-\Lambda+\bar{F}_K\left(-\frac{\ddot{a}}{a}\alpha-2H^{2}\alpha\right)+\frac{1}{2}a^2M^2F(\bar{K})- \dot{\bar{F}}_Ka\dot{a}\alpha\\
\end{align}
and the full (inhomogeneous) equations with the homogeneous parts subtracted are given by
\begin{align}
-2\frac{\nabla^2 V_N}{c^2a^2}&=\frac{8\pi G \bar{\rho} \delta}{c^2}-\frac{M^{2}}{2}\left(F-F(\bar{K})\right)+\frac{c_1}{c^2a^2}\left(U_{N,i} F_K\right)_{,i}\nn\\
&+3\frac{\alpha}{c^2}H^{2}\left(F_K-\bar{F}_K \right)+\frac{1}{c^2}U_NM^2F\hspace{1cm}\\
-\frac{1}{2ac^3}\left(4\dot{a}U_{N,i}+4a\dot{V}_{N,i}-\nabla^2B_{i} \right)&=-\frac{8\pi Ga \rho v_i}{c^3}+c_1F_K\frac{U_{N,i}}{c^3}H+\frac{c_1}{c^3}\dot{F}_K U_{N,i}-\frac{aB_i}{2c^3}M^2F\nonumber\\
&+\frac{(c_1-c_3)}{2c^3}\bigg(F_K\bigg(\left(\beta^{,l}_i-\beta^l_{,i} \right)+\frac{1}{a} \left(B^l_{,i}-B^{,l}_i \right)\bigg)\bigg)_{,l}\nn\\
&+\frac{F_K c_1\dot{U}_{N,i}}{c^3}\\
\frac{1}{c^2}\delta_{ij}\nabla^2(V_N-U_N) -\frac{1}{c^2}\left(V_N-U_N\right)_{,ij}&=\frac{a^2}{c^2}V_N\delta_{ij}M^2F\nonumber+\frac{a^{2}}{c^2}\delta_{ij}\bigg((F_K-\bar{F}_K)\left(-\frac{\ddot{a}}{a}\alpha-2H^2\alpha\right)\nn\\
&+\frac{M^{2}}{2}\left(F-F(\bar{K})\right)- \alpha(\dot{F}_K-\dot{\bar{F}}_K)H\bigg)\text{.}
\end{align}
The spatial part of the vector field equation is
\begin{align}
0 &=2\frac{c_1 U_{N,j}}{c^3}\dot{F}_K+F_{K,i}\bigg(2\delta^i_j\frac{\alpha}{c}H+2\delta^i_j\frac{\alpha U_N}{c^3}H+\frac{2\alpha\delta^i_j\dot{V}_N}{c^3}+2\delta^i_j\frac{c_2\beta^k_{,k}}{c^3}+2\frac{c_1\beta^{,i}_{j}}{c^3}+2\frac{c_3\beta^{i}_{,j}}{c^3}\nn\\
&+\frac{1}{ac^3}\bigg(c_1-c_3\bigg)\left(B^i_{,j}-B^{,i}_{j} \right)\bigg)+F_K\bigg(2\frac{c_1\dot{U}_{N,j}}{c^3}+2(\alpha+c_1)\frac{U_{N,j}}{c^3}H+\frac{2\alpha \dot{V}_{N,j}}{c^3}\nn\\
&+\partial_i\bigg(2\delta^i_j\frac{c_2\beta^k_{,k}}{c^3}+2\frac{c_1\beta^{,i}_{j}}{c^3}+2\frac{c_3\beta^{i}_{,j}}{c^3}+\frac{1}{ac^3}\left(c_1-c_3\right)\left(B^i_{,j}-B^{,i}_{j} \right) \bigg)\bigg)\nonumber
\end{align}
We can re-write the inhomogeneous equations in terms of the re-summed potentials, split the spatial perturbation to the vector field into its scalar and vector parts as $\beta^i\rightarrow \partial_i\xi^S+\xi^{i}$, and simplify the equations to get

\begin{align}
\label{eqn_newtonianresummed}
2\frac{\nabla^2 \psi_P}{c^2a^2}&=\frac{8\pi G \bar{\rho} \delta}{c^2}-\frac{M^{2}}{2}\left(F-F(\bar{K})\right)-\frac{c_1}{c^2a^2}\left(\phi_{P,i} F_K\right)_{,i}\nn\\
&+3\frac{\alpha}{c^2}H^{2}\left(F_K-\bar{F}_K \right)-\frac{1}{c^2}\phi_PM^2F\\
\frac{1}{2ac^3}\left(4\dot{a}\phi_{P,i}+4a\dot{\psi}_{P,i}+\nabla^2\omega_{i} \right)&=-\frac{8\pi Ga \rho v_i}{c^3}-c_1F_K\frac{\phi_{P,i}\dot{a}}{c^3a}-\frac{c_1}{c^3}\dot{F}_K \phi_{P,i}-\frac{a\omega_i}{2c^3}M^2F\nonumber\\
&+\frac{(c_1-c_3)}{2c^3}\bigg(F_K\bigg(\left(\xi^{,l}_i-\xi^l_{,i} \right)+\frac{1}{a} \left(\omega^l_{,i}-\omega^{,l}_i \right)\bigg)\bigg)_{,l}\nn\\
&-\frac{F_K c_1\dot{\phi}_{P,i}}{c^3}\\
-\frac{1}{c^2}\delta_{ij}\nabla^2(\psi_P-\phi_P) +\frac{1}{c^2}\left(\psi_P-\phi_P\right)_{,ij}&=-\frac{a^2}{c^2}\psi_P\delta_{ij}M^2F+\delta_{ij}\frac{1}{c^2}\bigg((F_K-\bar{F}_K)\left(-a\ddot{a}\alpha-2\dot{a}^2\alpha\right)\nn\\
&+\frac{a^{2} M^{2}}{2}\left(F-F(\bar{K})\right)- \alpha(\dot{F}_K-\dot{\bar{F}}_K)a\dot{a} \bigg)
\end{align}
and

\begin{align}
0&=F_{K,i}\bigg(2\delta^i_j\frac{\alpha}{c}H-2\delta^i_j\frac{\alpha\phi_P}{c^3}H-\frac{2\alpha\delta^i_j\dot{\psi}_P}{c^3}+2\delta^i_j\frac{c_2\xi^{S,k}_{,k}}{c^3}+2\frac{c_1(\xi^{S,i}_{,j}+\xi^{,i}_{j})}{c^3}+2\frac{c_3(\xi^{S,i}_{,j}+\xi^{i}_{,j})}{c^3}\nn\\
&+\frac{1}{ac^3}\left(c_1-c_3\right)\left(\omega^i_{,j}-\omega^{,i}_{j} \right)\bigg)-2\frac{c_1 \phi_{P,j}}{c^3}\dot{F}_K +F_K\bigg(-2\frac{c_1\dot{\phi}_{P,j}}{c^3}-2(\alpha+c_1)\frac{\phi_{P,j}}{c^3}H-\frac{2\alpha \dot{\psi}_{P,j}}{c^3}\nn\\
&+\partial_i\bigg(2\delta^i_j\frac{c_2\xi^{S,k}_{,k}}{c^3}+2\frac{c_1(\xi^{S,i}_{,j}+\xi^{,i}_{j})}{c^3}+2\frac{c_3\xi^{S,i}_{,j}}{c^3}-\frac{1}{ac^3}\left(c_1-c_3\right)\omega^{,i}_{j}  \bigg)\bigg)\text{.}
\end{align}
The scalar $K$ re-written in terms of the same variables is given by
\begin{equation}
\label{eqn_kresummed}
K=3\frac{\alpha}{M^2c^2}H^{2}+2\frac{\alpha\beta^i_{,i}}{M^2 c^4}H-c_1\frac{\phi_{P,k}\phi^{,k}_P}{M^2a^2c^4}-6\frac{\alpha\phi_{P}}{M^2 c^4}H^{2}-\frac{6\alpha\dot{\psi}_P}{M^2c^4}H
\end{equation}
Equations (\ref{eqn_newtonianresummed})-(\ref{eqn_kresummed}) represent the Newtonian limit of the equations governing cosmological structure formation, and apply regardless of the size of the density contrast. They can be compared to both the cosmological perturbation theory equations and galaxy-scale weak-field equations to understand the phenomenology of these theories and how the different regimes relate to each other. We will use these equations with the process laid out earlier to derive equations that govern cosmological structure formation on all scales.

\subsection{All scales equations}
\label{sec_allscales}
The leading order $\frac{1}{c}$ equations and linear perturbation equations (the latter are in appendix \ref{sec_linear}) can be combined to give

\begin{align}
2\frac{\nabla^2 \psi_P}{c^2a^2}&=\frac{8\pi G \bar{\rho} \delta}{c^2}-\frac{M^{2}}{2}\left(F-F(\bar{K})\right)-\frac{c_1}{c^2a^2}\left(\phi_{P,i} F_K\right)_{,i}\nn\\
&+3\frac{\alpha}{c^2}H^{2}\left(F_K-\bar{F}_K \right)-\frac{1}{c^2}\phi_PM^2F+\frac{1}{c^4}{\cal LS}_{A}\label{eqn_LSG00}\\
\frac{1}{2c^3}\left(4H\phi_{P,i}+4\dot{\psi}_{P,i}+\frac{1}{a}\nabla^2\omega_{i} \right)&=-\frac{8\pi Ga \rho v_i}{c^3}-c_1F_K\frac{\phi_{P,i}}{c^3}H-\frac{c_1}{c^3}\dot{F}_K \phi_{P,i}-\frac{a\omega_i}{2c^3}M^2F\nonumber\\
&+\frac{(c_1-c_3)}{2c^3}\bigg(F_K\bigg(\left(\xi^{,l}_i-\xi^l_{,i} \right)+\frac{1}{a} \left(\omega^l_{,i}-\omega^{,l}_i \right)\bigg)\bigg)_{,l}\nn\\
&-\frac{F_K c_1\dot{\phi}_{P,i}}{c^3}+\frac{1}{c^5}{\cal LS}_{B}\label{eqn_LSG0i}\\
-\frac{1}{c^2}\delta_{ij}\nabla^2(\psi_P-\phi_P) +\frac{1}{c^2}\left(\psi_P-\phi_P\right)_{,ij}&=-\frac{a^2}{c^2}\psi_P\delta_{ij}M^2F+\delta_{ij}\frac{a^{2}}{c^2}\bigg((F_K-\bar{F}_K)\left(-\frac{\ddot{a}}{a}\alpha-2H^{2}\alpha\right)\nn\\
&+\frac{M^{2}}{2}\left(F-F(\bar{K})\right)- (\dot{F}_K-\dot{\bar{F}}_K)H\alpha \bigg)+\frac{1}{c^4}{\cal LS}_{C}
\end{align}
and

\begin{align}
0 &=F_{K,i}\bigg(2\delta^i_j\frac{\alpha}{c}H-2\delta^i_j\frac{\alpha\dot{a}\phi_P}{c^3a}-\frac{2\alpha\delta^i_j\dot{\psi}_P}{c^3}+2\delta^i_j\frac{c_2\xi^{S,k}_{,k}}{c^3}+2\frac{c_1(\xi^{S,i}_{,j}+\xi^{,i}_{j})}{c^3}+2\frac{c_3(\xi^{S,i}_{,j}+\xi^{i}_{,j})}{c^3}\nn\\
&+\frac{1}{ac^3}\left(c_1-c_3\right)\left(\omega^i_{,j}-\omega^{,i}_{j} \right)\bigg)+F_K\bigg(-2\frac{c_1\dot{\phi}_{P,j}}{c^3}-2(\alpha+c_1)\frac{\phi_{P,j}}{c^3}H-\frac{2\alpha \dot{\psi}_{P,j}}{c^3}\bigg)\nn\\
&+F_{K}\partial_i\bigg(2\delta^i_j\frac{c_2\xi^{S,k}_{,k}}{c^3}+2\frac{c_1(\xi^{S,i}_{,j}+\xi^{,i}_{j})}{c^3}+2\frac{c_3\xi^{S,i}_{,j}}{c^3}-\frac{1}{ac^3}\left(c_1-c_3\right)\omega^{,i}_{j}  \bigg)-2\frac{c_1 \phi_{P,j}}{c^3}\dot{F}_K+\frac{1}{c^5}{\cal LS}_{D}\text{.}
\end{align}
We have denoted by ${\cal LS}$ the large scale terms that only apply on scales close to the horizon, the explicit form of which are given in Appendix \ref{app_large_scale_terms}. These terms can usually be neglected in N-body simulations (but do not need to be), see discussion in \cite{theorypaper}. Their inclusion here is still important as it ensures that we have a single set of equations that describe structure formation on all scales, containing the same variables and with the freedom in the theory chosen consistently. In other words, when examining the possibility of cosmological MONDian behaviour in section \ref{sec_discussion}, we can ensure that the behaviour in the different cosmological regimes are all consistent with each other and realisable in the same universe, with the same choice of $F(K)$ (including any conditions on $K$ for different branches of $F(K)$ to be realised).

\subsection{Consistently expanding $M^2$ and $F(K)$}
\label{sec_allscales2}
We now consider the issue of $M^2$ in the expansion in powers of $c$, and how this relates to expanding $F(K)$ and $F_K$. By construction, $K$ is dimensionless. However $M^2$ is not, and dimensionful constants typically need to be considered carefully in $\frac{1}{c}$ type expansions (see e.g. \cite{PPNV,theorypaper}). For the original linear EA theory $M^2K$, it makes no difference to the leading order equations which power of $c$ is assigned to $M$, however different choices will result in difference outcomes for the more general case with $F(K)$. The following considerations and observations will be used to guide our choice
\begin{itemize}
\item The power of $c$ attached to the leading order of $F_K$ will differ from that of the leading order of $F(K)$ by the power of $c$ attached to the leading order of $K$.
\item It makes little sense for  $M^2F(K)$ and $F_K$ to contribute at lower order than the standard GR and matter terms, thus requiring $M^2F(K)$ to be of order $c^{-2}$ or smaller and $F_K$ to be of order $c^{0}$ or smaller.
\item Considering power laws, $F(K)\propto K^n$, the only power of $c$ such that $M\rightarrow M_* c^m$ doesn't violate the conditions in the previous point for any power law $n$ is $m=-1$.
\item From the vector field equation, if the leading order background and inhomogeneous parts of $F_K$ are at the same order in $c$, the equation simplifies to $\delta F_{K,i}=0$; otherwise it should have a power of c at least two lower.
\item For \textit{any} power of $c$ assigned to $M$, if $\bar{K}=0$ then the choice of the functional form $F(K)$ is much more constrained in order to deliver sensible results. For example, a power law $F(K)=AK^n$ is problematic for the $\frac{1}{c}$ expansion if $n<0$. The same is probably true for a Taylor expansion of $F(K)$ and $F_K$ as typically performed in perturbation theory: if $\bar{K}=0$ then this expansion is problematic.
\end{itemize}
Given all of these, the choice $M^2\rightarrow\frac{M^2_*}{c^2}$ appears the most sensible, at least as long as there is a background for $K$ (i.e. $\alpha \neq0$). With this choice, the leading order part of $K$ purely homogeneous (i.e. the leading order $\bar{K}$ is $c^2$ larger than the leading order of $\delta K$), which makes sense in terms of expanding around a background and the dimensionless nature of $K$. It also makes the outcome more similar to a perturbative expansion, so many of the terms that could exist in principle in the non-linear regime are not present at leading order. The contributions at leading order become for a powerlaw ($K^n$)
\begin{eqnarray}
&&M^2F(\bar{K})\propto c^{-2}\text{;  } M^2F-M^2F(\bar{K})\propto c^{-4}\nonumber\\
&&\bar{F}_K\propto c^0\text{;  } F_K-\bar{F}_K \propto c^{-2} \nonumber
\end{eqnarray}
 i.e. the homogeneous parts contribute at exactly the expected order for all power laws for $F(K)$, and the inhomogeneous parts won't contribute at leading order. There are few explicit forms for $F(K)$ beyond power laws in the literature that we can use to test our choices further. One example is \cite{gealogfunc}, which uses $F(K)\sim\sqrt{K}+\sqrt{K}\ln{K}$. For the choices we have made, this function also gives a sensible expansion as long as $\alpha \neq0$.
%
With this choice, the equations become
\begin{align}
2\frac{\nabla^2 \psi_P}{c^2a^2}&=\frac{8\pi G \bar{\rho} \delta}{c^2}-\frac{c_1}{c^2a^2}\bar{F}_K\left(\phi_{P,i} \right)_{,i}+\frac{1}{c^4}{\cal LS}_{A2}\label{eqn_allscales2}\\
\frac{1}{2c^3}\left(4H\phi_{P,i}+4\dot{\psi}_{P,i}+\frac{1}{a}\nabla^2\omega_{i} \right)&=-\frac{8\pi Ga \rho v_i}{c^3}-c_1\bar{F}_K\frac{\phi_{P,i}}{c^3}H-\frac{c_1}{c^3}\dot{\bar{F}}_K \phi_{P,i}-\frac{\bar{F}_K c_1\dot{\phi}_{P,i}}{c^3}\nonumber\\
&+\frac{(c_1-c_3)}{2c^3}\bar{F}_K\bigg(\xi^{,l}_i-\frac{1}{a}\omega^{,l}_i \bigg)_{,l}+\frac{1}{c^5}{\cal LS}_{B2}\label{eqn_LSG0i_2}\\
-\frac{1}{c^2}\delta_{ij}\nabla^2(\psi_P-\phi_P) +\left(\psi_P-\phi_P\right)_{,ij}&=\frac{1}{c^4}{\cal LS}_{C2}
\end{align}
and

\begin{align}
0 &=\bar{F}_K\bigg(-2\frac{c_1\dot{\phi}_{P,j}}{c^3}-2(\alpha+c_1)\frac{\phi_{P,j}\dot{a}}{c^3a}-\frac{2\alpha \dot{\psi}_{P,j}}{c^3}\bigg)+F_{K}\partial_i\bigg(2\delta^i_j\frac{c_2\xi^{S,k}_{,k}}{c^3}+2\frac{c_1(\xi^{S,i}_{,j}+\xi^{,i}_{j})}{c^3}\nn\\
&+2\frac{c_3\xi^{S,i}_{,j}}{c^3}-\frac{1}{ac^3}\left(c_1-c_3\right)\omega^{,i}_{j}  \bigg)-2\frac{c_1 \phi_{P,j}}{c^3}\dot{\bar{F}}_K+2F_{K,j}\frac{\alpha}{c}H+\frac{1}{c^5}{\cal LS}_{D2}\text{,}
\label{eqn_allscales2_2}
\end{align}
where the large scale terms are denoted by ${\cal LS}$ and their explicit form is given in Appendix \ref{app_large_scale_terms}.

Note that not all of the terms that get removed from the leading order equations as a result of applying this choice for $M$ appear in the revised ``large scale terms'', since some of the removed terms no longer contribute at leading order in either limit, and thus they can be removed entirely. We also note that when examining MOND phenomenology in section \ref{sec_discussion} the unusual properties of the specific $F(K)$ that is used necessitates keeping some of the sub-leading terms that have been neglected in deriving this second set of equations. More generally, if one is concerned about the properties of a particular $F(K)$ then one can always insert $M^2\rightarrow\frac{M^2_*}{c^2}$ into the first set of all-scales equations and check whether any additional terms should be kept.

Taking the vector part of the spatial vector field and $G_{0i}$ equations (and ignoring the large scale terms) we find
\begin{align}
\label{eqn_vectest}
\nabla^2\xi_i&=\frac{(c_1-c_3)}{2ac_1}\nabla^2\omega_i\\
 \label{eqn_vector} \frac{1}{2ac^3}\nabla^2\omega_i\left(1+\bar{F}_K(c_1-c_3)-\bar{F}_K \frac{(c_1-c_3)^2}{2c_1} \right)&=-\frac{8\pi Ga \rho v_i}{c^3}\bigg\rvert_{V}\text{,}
\end{align}
which will be used in the next subsection.

\subsection{Conditions for the derivation to hold}
\label{sec_conditions}
Here we paraphrase section V.A of \cite{theorypaper} to formally lay out the conditions that are required for this derivation to be valid, and thus have been implicitly assumed above. Since the quantitative tests require detailed numerical simulations, we do not carry the tests out here, however we lay out these conditions to give a full picture of how the method works conceptually and how to test the assumptions that underlie the method. These assumptions are often explicitly or implicitly in other calculations carried out in modified gravity cosmologies (for example when using N-body simulations). 
 
As noted earlier, we don't require the first condition in section V.A of \cite{theorypaper} that only scalar fluctuations are important for structure formation, so the remaining requirements are equivalent to the two limits (linear perturbation and Newtonian) are valid on the largest and smallest scales, and every relevant scale is well-described by one (or both) of the two limits, i.e. there is no intermediate regime (see figure \ref{fig_scales}) where neither of the limits apply and there is at least a small region of overlap where both limits apply. As such, these requirements can be written more concretely as
\begin{enumerate}
\item A weak field metric is appropriate on all cosmological scales.
\item Check for the existence of a scale $k_*$, which is between the horizon scale and the scale at which the density fluctuations become non-linear, such that on length scales below the length scale corresponding to $k_*$ the only terms that contribute significantly in the linear perturbation equations are the same terms that are present in the linearised Newtonian equations.
\item Calculate the metric vector potential $\vec{\omega}$ from GEA N-body simulations using equation \ref{eqn_vectest}, and check that this is small enough on all non-linear scales for the Newtonian approximation to be valid on all of these scales.
\end{enumerate}

The first of these is assumed to be axiomatic in most cosmologies examined in the literature (including $\Lambda$CDM and modified gravity cosmologies) and there is no evidence to the contrary (see e.g.  the discussion in \cite{theorypaper}), so we do not discuss it further. The second and third conditions on this list are quantitative tests of the validity of this derivation that require numerically solving the equations.

The second test requires using a modified cosmological Boltzmann code that includes the GEA modifications to the linear perturbation equations, and in practice this is equivalent to testing the quasi-static approximation (see discussion in \cite{theorypaper}). The quasi-static approximation is a fairly standard approximation in modified gravity, and is typically assumed to hold in GEA theories (e.g. \cite{1711.09893}), so we assume the same here and leave an in-depth examination of this to future work.

The hardest condition to test is the third one, namely that the Newtonian limit of the gravitational equations is a good approximation on all scales below the scale at which density fluctuations become non-linear. This condition has been checked explicitly for GR+$\Lambda$CDM \cite{Bruni:2013mua,longerpaper} and $f(R)$ \cite{postffr}, and is implicitly assumed to be true whenever N-body simulations are run for a particular cosmology, so it can be seen as a formal check of the working assumptions under which simulations are run, irrespective of the issue we consider in this paper of how to consistently describe structure formation on all scales. This test requires N-body simulations for the cosmology in question, which can now be run for GEA theories using the equations laid out earlier.

\section{Discussion and phenomenology of the equations}
\label{sec_discussion}
Equations (\ref{eqn_allscales2})-(\ref{eqn_allscales2_2}) are the GEA equivalent to the $\Lambda$CDM+GR equations ($3.20$) in section IV of \cite{theorypaper}. These gravitational equations should be combined with the matter equations (3.20c) and (3.20d) from \cite{theorypaper} (repeated in appendix \ref{app_matter} for completeness) in order to provide a complete and unified set of equations for evolving the density fluctuations in the universe on any scale in a generalised Einstein-Aether cosmology. Here we examine these equations and their consequences.

One key point of these equations is that since cosmological structure formation becomes non-linear in the matter dominated era, and by construction the all-scales equations smoothly connect to the perturbation theory equations in the matter dominated era, then the combination of the equations here (describing the matter dominated universe onwards) and the (well-studied) cosmological perturbation theory for GEA theories comprises a complete and coherent description of GEA cosmologies. To the authors' knowledge this is the first complete and coherent cosmology of a relativistic extension of MOND covering all necessary scales and regimes. These equations also contain the source equation for the vector potential in the metric, which can be used to check the validity of the Newtonian limit on all non-linear scales \cite{Bruni:2013mua,longerpaper,theorypaper} and can generate novel lensing signals, so it is a useful equation to have derived for GEA cosmologies irrespective of its uses in the context  in this work.

We note that in these equations the same $F(K)$ (with the same branches and conditions for the branches to be realised) is present on all cosmological scales, and therefore the chosen function will govern all of the cosmological dynamics, including the background expansion, large (linear) scales and non-linear scales where the density contrast is large. For consistency of the theory and the cosmology, this $F(K)$ cannot be chosen differently in these different regimes (unless the branch conditions are such that a different part of $F(K)$ applies in each regime, which is unlikely numerically). As a result, we can see that introducing MONDian behaviour on non-linear scales is likely to also introduce MONDian behaviour into the cosmological perturbation equations (as long as the numerical values of $a_0$ and the gravitational potential are such that the perturbations are in the MONDian regime, or at least in transition regime between the Newtonian and MONDian branches). In other words these two regimes are intertwined and one cannot arbitrarily put MONDian behaviour into only one of the two regimes. We will see this in more detail in subsection \ref{sec_alphazero}.

In the weak field limit that is present in galaxies, deep MONDian behaviour arises using $\lim_{|\nabla \Phi| \ll a_0} F(K)=-\frac{2}{c_1}K+BK^{3/2}$  \cite{0607411}. When we apply this $F(K)$ to our cosmological equations, we find that in general MOND behaviour does not arise, due to the presence of the background terms in $K$, so we examine this choice of $F(K)$ separately for the two cases and $\alpha\neq0$ and $\alpha=0$. We start by assuming we are in the deep MOND regime in each case, and then later comment on choosing $F(K)$ for the Newtonian branch and the criteria determining the different branches for each case.

\subsection{$\alpha\neq0$}
Using $K$ as in equation (\ref{eqn_kresummed}) and $F(K)=-\frac{2}{c_1}K+BK^{3/2}$, we find that the leading order (in $\frac{1}{c}$) $G_{00}$ equation is given by
\begin{equation}
\frac{\nabla^2\phi_P}{c^2}=\frac{8\pi Ga^2 \bar{\rho} \delta}{c^2}\frac{2a M_*}{3c_1 B \dot{a} \sqrt{3\alpha}}
\end{equation}
The contributions that are usually responsible for delivering MONDian behaviour are higher order $\left(c^{-3}\right)$ in this case, so they do not contribute at leading order. There are no issues performing the perturbative expansion in this case, and the perturbative equations match the $\frac{1}{c}$ equations in the overlap regime, so the all scales equations follow from substituting the appropriate $K$ and $F_K$ into the equations in section \ref{sec_allscales2}. We do not make this substitution here for brevity.
 
It is often taken that $M_*\sim H_0$; in this case there is an order unity change to Newton's constant in the cosmological Poisson equation, with no scale dependence and only simple time dependence. Thus, GEA theories would not manifest any cosmological MONDian behaviour if the background expansion is modified from the GR Friedmann equations. These theories might still have MONDian behaviour on galaxy scales below where the FLRW metric applies. As MONDian behaviour does not arise for this case, we do not look further into the issue of defining different branches or choosing different $F(K)$s in different branches.

\subsection{$\alpha=0$}
\label{sec_alphazero}
We now consider the case $\alpha=0$, such that $K=-c_1\frac{\phi_{P,k}\phi^{,k}_P}{M_*^2a^2c^2}$. This is a somewhat unusual $K$ in that $\bar{K}=0$, but $F_K$ contains a constant (and thus effectively homogeneous and not perturbatively small) part for any $K$. As such, a Taylor expansion of $F(K)$\footnote{I.e. expanding as $F(K)\rightarrow F(\bar{K})+F_K(\bar{K})\delta K$ and $F_K\rightarrow F_{K}(\bar{K})+F_{KK} (\bar{K})\delta K$.} (as sometimes used perturbatively) fails. Truncating the $G_{00}$ equation at order $c^{-2}$ gives $\frac{8\pi G \bar{\rho} \delta}{c^2}=0$, so the leading order equation (in the $c^{-1}$ expansion) must be expanded to include additional terms of order $\frac{1}{c^3}$ or higher. The $K^{\frac{3}{2}}$ part of $F(K)$ does not contribute at order $c^{-2}$, but it does contribute at $c^{-3}$ order through the term $-\frac{c_1}{c^2a^2}\left(\phi_{P,i} F_K\right)_{,i}$ in equation (\ref{eqn_LSG00}). The other beyond-leading-order terms arise at $c^{-4}$ or higher order, or are of order $\frac{1}{c^3}$ but occur with a prefactor $\alpha$, so disappear for this case, leaving
\begin{equation}
\frac{3Bc_1}{2M_* a c^3}\bigg(\phi_{P,i}\sqrt{-c_1 \phi_{P,k}\phi^{,k}_P} \bigg)_{,i}=\frac{8\pi Ga^2 \bar{\rho} \delta}{c^2} \text{.}
\end{equation}
 As such, the only terms at this order are exactly the terms that give rise to a MONDian Poisson equation as in the galactic weak field case, therefore in this case it is possible to get MONDian behaviour on cosmological scales. 
 
The unusual nature of the choice of $F(K)$ is what makes the derivation of MOND behaviour in this way in a relativistic extension of MOND somewhat strange, in that $F_K$ cancels the leading order term from GR, and replaces it with a term that is nominally not leading order, but is actually the dominant term that remains after cancelling the usual leading order terms. An analogous cancellation occurs in the weak field galactic MOND case. Allowing for this case in the general $G_{00}$ all-scales equation requires only a very minor adjustment (reverting one of the terms from equation (\ref{eqn_allscales2}) to its form in equation (\ref{eqn_LSG00})),
\begin{equation}
2\frac{\nabla^2 \psi_P}{c^2a^2}=\frac{8\pi G \bar{\rho} \delta}{c^2}-\frac{c_1}{c^2a^2}\left(\phi_{P,i} F_K\right)_{,i}+\frac{1}{c^4}{\cal LS}_{A2}\text{.}
\end{equation}
This equation replaces equation (\ref{eqn_allscales2}) when it is desired to include GEA theories with cosmological MONDian behaviour.
 
One of the advantages of the all-scales equations we have constructed is that the leading order (in the $c^{-1}$ expansion) parts of these equations cover the quasi-static linear and non-linear regimes consistently and simultaneously. As such, we can see how this MONDian behaviour manifests on scales that are well below the horizon where the density contrast is small: this is what is usually referred to as the (perturbative) quasi-static regime. If one naively applied the quasi static approximation and linear perturbation theory in this regime, then one would have a nonsenical leading order equation and no MONDian behaviour, despite the Newtonian limit of the (non-linear) equations suggesting that such a MONDian term should be present and dominant. The combination of the all-scales equations and the tests to determine their validity gives a way out of this problem. Firstly one can run simulations with the leading order $\frac{1}{c}$ equations and determine what scale these are valid up to (using the vector potential check described earlier) and what size the density contrast is on those scales. If the $\frac{1}{c}$ equations are valid up to a scale where the density contrast is small (linear), then the problem is the linearisation in perturbation theory preventing the MONDian term. One can then use the linearised all-scales equations in a Boltzmann code and check that the quasi static approximation is valid to a larger scale than the non-linear scale (from the perspective of perturbation theory). If this second test is passed then the all-scales equations are valid, despite the problem with a naive application of perturbation theory. If either test is failed, then this means the cosmology in question, unlike $\Lambda$CDM, has a range of scales where neither perturbation theory nor the Newtonian limit is valid. The authors are not aware of any tools for computing cosmological structure formation in any cosmology that has such a regime. Relatedly, since the all-scales equations do not apply to a radiation-dominated cosmology, it remains unclear if and how cosmological MONDian behaviour might arise in the perturbation equations during the radiation dominated era.

As deep MOND behaviour can manifest for this case, we now consider how a Newtonian branch could behave, and what the conditions for being in different branches would be.\footnote{We assume that these branch conditions are well defined and do not worry about for example whether different observers see a different $\nabla \Phi$, or the issue of whether the background equations should use the deep MOND branch of $F(K)$ due to having no peculiar acceleration by definition; this latter issue is not so important anyway given our result that MONDian behaviour only arises when $\bar{K}=0$.} The condition to be in the deep MONDian branch of equation (\ref{eqn_cosmomond}) is $\frac{|\nabla \phi_P|}{\gamma(a)a_0}\ll 1$. By comparing the GEA Poisson equation in this case to the MOND Poisson equation we can identify
\begin{eqnarray}
&&\gamma(a) a_0=\frac{4ac^4M}{3Bc_1\sqrt{-c_1}}\\
&&a_0=\frac{4c^4M}{3Bc_1\sqrt{-c_1}}\\
&&\gamma(a)=a \text{,}
\end{eqnarray}
where this identification is independent of any power of $c$ assigned to $M$. We have observed from the first equation that for GEA theories manifesting cosmological MOND, $\gamma(a)=a$, however we leave $\gamma(a)$ in the equations for the rest of this subsection to ease comparison with the deep MONDian branch of equation (\ref{eqn_cosmomond}). Interestingly, this form of $\gamma(a)$ means that when converted to a physical derivative, the condition to be in the deep MONDian branch, and thus the effective $a_0$, does not vary with time. This is similar to the behaviour found in a different relativistic extension of MOND in \cite{0802.1526}. We can substitute into the expression for $K$ and define $K_*=\frac{16c^4}{9B^2c^2_1}$ to relate $K$ to the definition of the deep MOND branch
\begin{equation}
\frac{K}{K_*}=\frac{|\nabla \phi|^2}{\gamma^2(a)a^2_0}\text{.}
\end{equation}
As such, the MONDian and Newtonian regimes are defined by how $K$ compares to the (constant) critical value $K_*$, with some explicit time dependence in the relationship. The MONDian branch and MONDian behaviour occurs when $0 < K\ll K_*$, and the Newtonian branch occurs when $K\gg K_*$. 

In principle, the function $F(K)$ can be chosen to have a different form in the Newtonian branch, however despite this the Newtonian branch is significantly constrained, as the term that modifies the Poisson equation is proportional to $\bar{F}_K\nabla^2 \phi_P$, which will not contribute for most $F(K)$ if $\alpha=0$. As such, requiring cosmological MOND behaviour in GEA theories means that the Newtonian branch in the quasi-static limit of perturbation theory is the GR Poisson equation, i.e. there is no modification to gravity and $G_\text{eff}=1$ in this branch, including in linear perturbation theory.

There is one final technical nuance to consider, which is whether the emergence of MONDian behaviour, the identification with $a_0$ and the conditions for the different branches are consistent with how we have derived our equations and the power of $c$ assigned to $M$. Assuming no power of $c$ assigned to $M$, the leading order term that we have neglected is
\begin{equation}
-\frac{a^2}{2}\left(MF(K)-MF(\bar{K}) \right)=-\frac{a^2}{2}\left(\frac{2\phi_{,k}\phi^{,k}}{a^2c^4}-\frac{Bc_1\sqrt{-c_1}}{Ma^3c^6}\biggl(\phi_{,k}\phi^{,k} \biggr)^{3/2} \right)\text{.}
\end{equation}
The second term is always sub-dominant to the MONDian term, as they both scale as $M^{-1}$ and the MONDian term has fewer powers of the potentials. The first term doesn't depend on how we expand $M$, whereas the MONDian term goes as $M^{-1}c^4$, so if no power of $c$ is attached to $M$ then in principle both of these terms contribute. For the MONDian term to dominate we need $M=M_*c^m$ with $m$ negative, i.e. $M$ should have some level of smallness associated to it. Since $M$ relates to $a_0$ then this is consistent with $a_0$ being small, and is all consistent with our choice of $m=-1$.

As an additional consistency check, we can also estimate the effect of the first term in the galactic weak-field regime by noting that it is akin to adding a term $\vec{\nabla} \Phi \cdot \vec{\nabla} \Phi$ on the left hand side of \ref{eqn_mond}. It can be shown that this modifies the solution $\Phi$ in the deep MOND branch at $a=1$ from $\partial_{r}\Phi = \sqrt{GMa_{0}}/r$ to $\partial_{r}\Phi = e^{-a_{0} r }\sqrt{GMa_{0}}/r$ so its effect is subdominant for $r\ll a_{0}$. Typical values for $a_{0}$ correspond to length scales of the order of the Hubble radius and so the first term is expected to be subdominant for the subhorizon scales of interest.

Since $\nabla \phi_P$ is free to vary substantially numerically just as in a usual MOND weak field expansion, the conditions for the branches of the MOND Poisson equation ($|\nabla \phi|>>\gamma(a)a_0$ and $|\nabla \phi|<<\gamma(a)a_0$) can both be realised without violating any assumptions in the derivation presented here (including the power of $c$ assigned to $M$). Note that in the discussion in this subsection, we do not comment on the scales and times in the universe where the different branch conditions would be realised numerically. Rather we are just showing what the branch conditions are, what the equations are in the limiting case of each branch, and that the branch conditions and equations are consistent across the different cosmological regimes (so e.g. the condition to be in the deep MOND regime is not different depending on whether the cosmological density contrast is non-linear or perturbative). We leave a full numerical solution of this model to future work, here we just note that the condition for the different branches is something that needs to be considered within a cosmological Boltzmann code if one is seeking to calculate predictions for a theory with MOND on cosmological scales.

\subsubsection{The $G_{0i}$ equation}
If we apply $\alpha=0$ and the same ansatz for $F(K)$ (in the deep MOND limit $0<K\ll K_*$) to the scalar part of the $G_{0i}$ equation, this equation behaves similarly to the Poisson equation: the leading order GR terms are again cancelled by the leading order parts of $F_K$. The scalar part of $G_{0i}$ is given by
\begin{equation}
\frac{1}{2c^3}\left(4H\phi_{P,i}+4\dot{\psi}_{P,i} \right)=-\frac{8\pi Ga \rho v_i}{c^3}-c_1F_K\frac{\phi_{P,i}}{c^3}H-\frac{c_1}{c^3}\dot{F}_K \phi_{P,i}-\frac{F_K c_1\dot{\phi}_{P,i}}{c^3}\text{,}
\end{equation}
where any terms that are not pure scalars are considered to have had their pure vector (divergenceless) parts subtracted. The leading order $F(K)$ term is  $F_K=-\frac{2}{c_1}$, which results in
\begin{equation}
\frac{2}{c^3}\left(H\phi_{P,i}+\dot{\psi}_{P,i} \right)=-\frac{8\pi Ga \rho v_i}{c^3}+2\frac{\phi_{P,i}}{c^3}H+\frac{2\dot{\phi}_{P,i}}{c^3}
\end{equation}
Since $\phi_P=\psi_P$ at leading order for any $F(K)$, this means all of the non-matter terms vanish. In other words, the MONDian function that is designed to cancel the leading order non-matter part in the GR Poisson equation, also cancels the leading order non-matter part in the GR $G_{0i}$ equation.

As for the Poisson equation we can go beyond leading order. The subleading terms that create MONDian behaviour in the Poisson equation will also be the terms of lowest remaining order here (these terms are of order $c^{-4}$, whereas the large scale B2 terms in equation (\ref{eqn_vectorLS}) are of order $c^{-5}$), so the scalar $G_{0i}$ equation in the deep MOND limit is given by 
\begin{equation}
\frac{c_1}{c^4}\bigg(\frac{3B}{2}\sqrt{\frac{-c_1\phi_{P,k}\phi^{,k}_P}{M^2_*a^2}}\big(H\phi_{P,i}+\dot{\phi}_{P,i} \big)+\frac{3B\phi_{P,i}}{2}\frac{\partial}{\partial t}\sqrt{\frac{-c_1\phi_{P,k}\phi^{,k}_P}{M^2_*a^2}} \bigg)\bigg\rvert_s=-\frac{8\pi Ga \rho v_i}{c^3}\bigg\rvert_s
\end{equation}
To the authors' knowledge, such an examination of the $G_{0i}$ Einstein equation in the deep MOND regime has not previously been carried out. It is interesting that a similar cancellation  occurs in both the Poisson equation and the $G_{0i}$ equation for the same choice of $F(K)$ and $\alpha=0$. We leave to future work an investigation of this potential coincidence, its possible physical significance, and whether it occurs in other relativistic extensions of MOND.

As for the $G_{00}$ equation, this case can be included in the all-scales $G_{0i}$ equation by simply returning a few of the terms that were dropped earlier, 
\begin{align}
\frac{1}{2c^3}\left(4H\phi_{P,i}+4\dot{\psi}_{P,i}+\frac{1}{a}\nabla^2\omega_{i} \right)&=-\frac{8\pi Ga \rho v_i}{c^3}-c_1 F_K\frac{\phi_{P,i}\dot{a}}{c^3a}-\frac{c_1}{c^3}\dot{F}_K \phi_{P,i}-\frac{F_K c_1\dot{\phi}_{P,i}}{c^3}\nonumber\\
&+\frac{(c_1-c_3)}{2c^3}\bar{F}_K\bigg(\xi^{,l}_i-\frac{1}{a}\omega^{,l}_i \bigg)_{,l}+\frac{1}{c^5}{\cal LS}_{B2} \text{,}
\end{align}
where again the explicit form of the large scale term ${\cal LS}_{B2}$ is given in Appendix \ref{app_large_scale_terms}. This equation replaces equation (\ref{eqn_LSG0i_2}) when it is desired to include GEA theories with cosmological MONDian behaviour.

\subsection{N-body simulations}
The all-scales equations derived above can be implemented in cosmological N-body simulations to give simulations that represent GEA cosmologies, with a consistent background expansion, and
where the inhomogeneities are evolved correctly no matter whether they are outside the horizon, around the horizon scale, well inside the horizon, or in the non-linear regime. In practice, for most cosmological simulations, particularly those with a smaller box size, the ``large scale'' terms can be neglected in the equations and this should make little difference to the output of the simulations. Initial conditions for these simulations can be calculated from a Boltzmann solver that implements GEA perturbation theory. Due to the smooth connection during the matter dominated era, these initial conditions and simulations will form a self-consistent and coherent complete cosmological picture for GEA theories. In particular, if models with $\alpha=0$ (and $F(K)$ as described above for MONDian behaviour) were run, these simulations would correspond to the first fully consistent N-body simulations for a relativistic extension of MOND where MONDian behaviour arises cosmologically.

We can also use these equations to examine how GEA cosmologies relate to cosmological MOND N-body simulations that have already been run \cite{0109016,0303222,0809.2899,1104.5040,1309.6094,1410.3844,1605.03192}. These works themselves acknowledge the complications of running consistent cosmological MOND simulations, and in particular include discussions about different choices and options regarding the background expansion history and the initial conditions. The point we wish to focus on here is a further issue, which was alluded to in \cite{1605.03192}: what are the actual equations governing the inhomogeneities in a consistent cosmology for a relativistic extension of MOND, and does this correspond to an implementation of MOND as carried out in N-body simulations that have been run so far.

As discussed earlier, when MOND is phrased as a modification to the Poisson equation, it is perhaps intuitive when one is used to GR to think that this can be applied to the Poisson equations that arise in different regimes, however these are not necessarily so similar in a modified gravity theory, and particularly a relativistic extension of MOND. This is borne out explicitly by the equations that we have derived above: for $\alpha\neq0$, although these models may still contain MOND on galactic scales, a cosmological N-body simulation should \textit{not }contain MONDian behaviour, even if the free function $F(K)$ in these theories is the same cosmologically as on galactic scales. So, despite having MOND on galactic scales these models do not relate to cosmological MOND N-body simulations.

For models with $\alpha=0$ however, there is a connection between MONDian behaviour on galactic and cosmological scales. In this case, we have shown that a Poisson equation with both MONDian and Newtonian branches exists, that the condition to be in either of these branches matches the usual MOND condition, and that both branches can be realised in principle. In particular, the Newtonian branch Poisson equation matches that in GR (for the vast majority of $F(K)$, i.e. unless $F(K)$ in the Newtonian branch is chosen very specifically to avoid this). In addition, we have shown that in these models  the background Friedmann equations are not modified from those in GR. Combined, these results mean that existing MOND N-body codes can be used to represent structure formation in these theories, for a particular choice of matter, as long as the simulation parameters for the matter (for the background and the particle content) are chosen consistently, as they would be in GR. It may be however that the initial conditions used in previous works do not match those required to interpret the output of these simulations as part of a consistent GEA cosmology; determining this is beyond the scope of the work here. As noted above, the equations in this paper can be implemented in a Boltzmann code and used to generate initial conditions that would be consistent with interpreting the output of the N-body codes as representative of structure formation in these GEA models containing cosmological MOND.

\subsection{Implications for GEA theories' ability to replace cold dark matter}
It appears that GEA theories that have cosmological MONDian behaviour are quite limited, in that the Newtonian branch of the quasi-static (linear and non-linear) Poisson equation and the Friedmann equations are all essentially restricted to have their GR form. Given the range of evidence for cold dark matter and the different scales and environments in which cold dark matter phenomenology appears, these theories are probably too restricted to make for good alternatives to $\Lambda$CDM without themselves adding additional matter species to the universe, although we do not investigate this issue in detail here.

However, we also see that it is not required for GEA theories to have cosmological MOND just because it arises on galactic scales. It may be that the GEA theories with only galactic MOND can keep the successes of the MOND paradigm without some of the problems (such as galaxy clusters), but determining this will require running cosmological simulations with the equations derived here and no cosmological MOND behaviour. The same situation may be true for other relativistic extensions of MOND. It may be that attempts to build a MOND+sterile neutrino paradigm (see e.g. \cite{0805.4014,1104.5040}) can benefit from the finding that MOND on cosmological scales is not required by having MOND on galactic scales; we leave a detailed examination of this prospect to future work.

\section{Conclusion}
\label{sec_conc}
To fairly and fully compare MONDian cosmologies against $\Lambda$CDM, we need to be able to examine cosmological structure formation on all scales in relativistic extensions of MOND, in a coherent and consistent fashion. In this paper we have laid out a method for deriving equations that govern the evolution of cosmological inhomogeneities on all scales in these theories. For any given theory, these equations allow consistent cosmological simulations to be run, and to examine when and how MONDian behaviour arises cosmologically. We have illustrated this method using the concrete example of GEA theories, resulting in the set of equations (\ref{eqn_allscales2})-(\ref{eqn_vector}) that govern the cosmological background and evolution of cosmological inhomogeneities on all scales in a unified manner with a single $F(K)$.

These equations show that MONDian behaviour does not necessarily arise cosmologically, even if the free function $F(K)$ in these theories is the same cosmologically as on galactic scales. Specifically, cosmological MONDian behaviour requires the model parameter $\alpha$ to be zero, and thus the background expansion history of the universe and the Newtonian branch of the cosmological Poisson equation should both be governed by the same equations as in GR. Although structure formation as carried out in existing MOND N-body simulations can correspond to these GEA theories, in practice these theories are quite limited, but it seems difficult to get cosmological MONDian behaviour without these limitations. It may be that the GEA theories with only galactic MOND can keep the successes of MOND without some of the problems (such as galaxy clusters), but determining this will require running cosmological simulations with the equations derived here. These results demonstrate the strengths of our method for deriving a single set of equations that govern cosmological structure formation on all scales.

We have also laid out the assumptions and checks required to validate our method for a given cosmology and commented on these for our case study; notably we derived the source term for the vector potential in GEA theories, equation (\ref{eqn_vector}), which can be extracted from N-body simulations and should be suitably small on all scales. If larger than in GR (but small enough to not jeopardise the $\frac{1}{c}$ expansion), this vector potential can also generate novel lensing phenomenology and thus provide a smoking gun for modified gravity behaviour on non-linear scales \cite{rotationest}.

The process laid out in this manuscript can be applied to any relativistic extension of MOND, thus paving the way for consistent cosmological simulations of these theories covering many decades in scale in order to fairly and fully compare them to the $\Lambda$CDM paradigm. It may also be the case that simply by deriving and analysing the equations for other relativistic extensions, as we have done here, the connection between some of these theories and galactic MONDian behaviour becomes clearer, potentially highlighting shortcomings in these theories without the need to run simulations. However, we caution against inferring what conclusions might be reached by such studies of a specific relativistic extension of MOND from the single study carried out here. It is clear from this work that the cosmological phenomenology of such theories is very rich, and trying to understand each of these theories consistently across all scales, including in which regimes MONDian behaviour can manifest, is not straightforward. 

In particular, one recent theory that it would be interesting to apply our method to is the AeST (Aether Scalar Tensor) theory \cite{2007.00082,2109.13287}. This theory satisfies the generic constraint from the CMB that the additional fields in these theories must behave similarly to the behaviour of cold dark matter in linear perturbation theory in the conditions that exist in the early universe \cite{gdmwtime,spergel}, but the extra fields in this theory can still behave substantially differently to cold dark matter in other regimes, potentially giving rise to MONDian behaviour (on galactic or cosmological scales) or cosmological structure formation that proceeds differently to GR+CDM structure formation. In this sense, whilst the theory can be argued to possess dark matter for the purposes of CMB calculations, it doesn't possess particle dark matter in the sense of the ``CDM'' in $\Lambda$CDM, or in the sense of a dark matter particle of the kind that would be detected in direct, indirect, or collider searches for dark matter. The method laid out in this paper provides a way to examine cosmological structure formation in this theory and determine in which regimes the behaviour is similar or different to GR with cold dark matter. Interestingly, in AeST theory the role of the cosmological dark matter overdensity involves the time derivative of a scalar field perturbation, which may require careful thought when defining a Newtonian limit.

\section*{Acknowledgements}
DBT Thanks Theo Anton for useful discussions. DBT acknowledges support from STFC Grant ST/T000341/1. This research is part of the project No. 2021/43/P/ST2/02141 co-funded by the Polish National Science Centre and the European Union Framework Programme for Research and Innovation Horizon 2020 under the Marie Sk\l{}odowska-Curie grant agreement No. 945339. This research has made use of NASA's Astrophysics Data System Bibliographic Services.

\appendix

\section{Linear perturbation equations}
\label{sec_linear}

In this appendix we present the linear perturbation equations written in terms of the resummed potentials.  Note that we have split things into their background and perturbative parts, i.e. $F_K\rightarrow\bar{F}_K+\delta F_K$ and $M^2F\rightarrow M^2\bar{F}+M^2\delta F$.\\
\paragraph{Homogeneous equations}
\begin{align}
0 &= \frac{1}{2}a^{2}(M^{2}\bar{F}-16\pi G\rho)+3a^{2}H^{2} - 3\alpha a^{2}H^{2}\bar{F}_K\\
0 &=  -3a^{2}\delta_{ij}H^{2}(1-\alpha \bar{F}_K) - \frac{1}{2}\delta_{ij}a^{2}(M^{2}\bar{F}+16\pi G P -2\dot{H}(-2+\alpha \bar{F}_K)-2\alpha H\partial_{T}(\bar{F}_K))
\end{align}

\paragraph{Time-Time component of the Einstein equations}
\begin{align}
0 &= \frac{1}{2}a^{2}M^{2}\delta F - 3\alpha a^{2}H^{2}\delta F_K  -8\pi G\delta\rho a^{2} + 2\partial_{i}\partial^{i}\psi_P+ c_{1} \bar{F}_K\partial_{i}\partial^{i}\phi_P  \nn\\
& +c_{1}a^2H\bar{F}_K \partial_{i}  \partial^{i}\xi^S -2\alpha a^2H \bar{F}_K \partial_{i}\partial^{i}\xi^S  + c_{1} a\bar{F}_K \partial_{i}\partial^{i}\partial_T(a\xi^S)  + a^{2}(-6H+6\alpha H\bar{F}_K)\dot{\psi}_P \nn\\
&+a^{2}(M^{2}\bar{F}-16\pi G \rho)\phi_P 
\end{align}
\paragraph{Space-Space components of the Einstein equations}
\begin{align}
0 &=  3\alpha a^{2}\delta_{ij}H^{2}\delta F_K - \frac{1}{2}\delta_{ij}a^{2}M^{2}\delta F + \delta_{ij}a^{2}\alpha\dot{H}\delta F_K+ \delta_{ij}a^{2}\alpha H\dot{\delta F_K}+ \delta_{ij}a^{2}(2-\alpha \bar{F}_K)\ddot{\psi}_P \nn\\
& -\partial_{i}\partial_{j}(\phi_P-\psi_P)+\delta_{ij}\partial_{k}\partial^{k}(\phi_P-\psi_P)+ a^{2}\delta_{ij}M^{2} \bar{F}\psi_P- 2aH \partial_{(i}\omega_{j)} -a\partial_{(i}\dot{\omega}_{j)} \nn\\
&+ \frac{1}{a}(c_{1}+c_{3})\partial_{T}(a^{3}\bar{F}_K \partial_{(i}\xi_{j)})+ (c_{1}+c_{3})\frac{1}{a}\partial_{T}(a^{3}\bar{F}_K \partial_{i}\partial_{j}\xi^S)- a^{2}\delta_{ij}(-2+\alpha \bar{F}_K)H \dot{\phi}_P\nn\\
& + a^2\delta_{ij}\alpha H\bar{F}_K \partial_{k} \partial^{k}\xi^S+\frac{c_{2}}{a}\delta_{ij}\partial_{T}(a^{3}\bar{F}_K \partial_{k}\partial^{k}\xi^S)   \nn\\
& + 2a^{2}\delta_{ij}(8\pi G P + 3H^{2}+\dot{H}(2-\alpha \bar{F}_K)-\alpha\frac{H}{a^{3}}\partial_{T}(a^{3} \bar{F}_K ))\psi_P+a^{2}\delta_{ij}(6H - \alpha\frac{1}{a^{6}}\partial_{T}(a^{6}\bar{F}_K))\dot{\psi}_P \nn\\
&+ 2a^{2}\delta_{ij} \phi_P(3H^{2}+\dot{H}(2-\alpha \bar{F}_K) - \alpha\frac{H}{a^{3}}\partial_{T}(a^{3}\bar{F}_K))
\end{align}
\paragraph{Time-Space components of the Einstein equations}
\begin{align}
&0 = 2\partial_{i}\dot{\psi}_P+8\pi G a v_i\rho +\frac{1}{2a}\partial_{j}\partial^{j}\omega_{i}+M^{2}F(\frac{1}{2}a\omega_{i}) + c_{1}\partial_{T}\bigg(\bar{F}_K \partial_{i}\phi_P\bigg) + (2H+c_{1}H\bar{F}_K)\partial_{i}\phi_P \nn\\
&-\frac{1}{2}(c_{1}-c_{3})\bar{F}_K\partial_{j}\bigg(  -\frac{1}{a}\partial^{j}\omega_{i}+\partial^{j}\xi_{i}\bigg) +a^2\bigg(2c_{1}H^{2}\bar{F}_K + (c_{1}-\alpha)\partial_{T}(H\bar{F}_K )\bigg)(\xi_{i}+\partial_{i}\xi^S)\nn\\
&+ c_{1}a\partial_{T}\bigg(\bar{F}_K (-\dot{\omega}_{i}+a\dot{\xi}_{i}+\dot{a}\xi_{i}+a \partial_{i}\dot{\xi}^S+\dot{a} \partial_{i} \xi^S)\bigg)+3c_{1}aH\bar{F}_K (-\dot{\omega}_{i}+a\dot{\xi}_{i}+\dot{a}\xi_{i}+a\partial_{i}\dot{\xi}^S+\dot{a}\partial_{i}\xi^S)\nn\\
&-\frac{1}{2}a\omega_{i}(-6H^{2}-4\dot{H}+16\pi G\rho+4c_{1}H^{2}\bar{F}_K+ 6\alpha H^{2}\bar{F}_K + 2c_{1}\partial_{T}(H\bar{F}_K))
\end{align}
\paragraph{Vector equation}
\begin{align}
0 &= -\frac{1}{a}(c_{1}-c_{3})\bar{F}_K\partial_{j} \partial^{j}\omega_{i}+ 2c_{1}\bar{F}_K\partial_{j} \partial^{j}\xi_{i}  -2\alpha\bar{F}_K \partial_{i} \dot{\psi}_P -2\alpha H\bar{F}_K \partial_{i} \phi_P -2c_{1}\frac{1}{a}\partial_{T}(a\partial_{i}\phi_P \bar{F}_K) \nn\\
&+2(c_{1}+c_2+c_{3})\bar{F}_K\partial_{j} \partial^{j}\partial_{i}\xi^S + \frac{2\alpha \dot{a}}{a}\partial_{i}\delta F_K -2c_{1}a\partial_{T}(F_{K}(-\dot{\omega}_{i}+a\dot{\xi}_{i}+\dot{a}\xi_{i}+a\partial_{i}\dot{\xi}^S+\dot{a}\partial_{i}\xi^S)) \nn\\
&- 6c_{1}aH \bar{F}_K (a\partial_{i}\dot{\xi}^S+\dot{a}\partial_{i}\xi^S+a\dot{\xi}_{i}+\dot{a}\xi_{i}-\dot{\omega}_{i})-a(2c_{1}-\alpha)\partial_{T}(H\bar{F}_K)(a\partial_{i}\xi^S+a\xi_{i}-\omega_{i})\nn\\
&  -4c_{1}aH^{2}\bar{F}_K (a\partial_{i}\xi^S+a\xi_{i}-\omega_{i})  
\end{align}

\section{Matter equations}
\label{app_matter}
These are the ``all scales'' matter equations from \cite{theorypaper}
\begin{align}
\frac{dv_i}{dt}&=-Hv_i+\frac{\phi_{P,i}}{a}\\
\frac{d\delta}{dt}&=-\frac{v_{i,i}}{a}\left(1+\delta \right)+\frac{3}{c^2}\frac{d\psi_P}{dt}
\end{align}

\section{Large scale terms}
\label{app_large_scale_terms}
\begin{align}
{\cal LS}_{A} &=-c_{1}H \bar{F}_K\partial_{i}  \partial^{i}\xi^S +2\alpha H \bar{F}_K \partial_{i}\partial^{i}\xi^S  -\frac{ c_{1}}{ a}\bar{F}_K \partial_{i} \partial^{i}\partial_T(a\xi^S)  + (6H-6\alpha H\bar{F}_K)\dot{\psi}_P\nn\\
&+16\pi G \rho\phi_P  
\end{align}
\begin{align}
{\cal LS}_{B}&= -a^2\bigg(2c_{1}H^{2}\bar{F}_K + (c_{1}-\alpha)\partial_{T}(H\bar{F}_K )\bigg)(\xi_{i}+\partial_{i}\xi^S)\nonumber\\
&+\frac{1}{2}a\omega_{i}(-6H^{2}-4\dot{H}+16\pi G\rho+4c_{1}H^{2}\bar{F}_K+ 6\alpha H^{2}\df + 2c_{1}\partial_{T}(H\bar{F}_K))\nn\\
&-c_{1}a\partial_{T}\bigg(\bar{F}_K(-\dot{\omega}_{i}+a\dot{\xi}_{i}+\dot{a}\xi_{i}+a \partial_{i}\dot{\xi}^S+\dot{a} \partial_{i} \xi^S)\bigg)\nn\\
&-3c_{1}aH\bar{F}_K (-\dot{\omega}_{i}+a\dot{\xi}_{i}+\dot{a}\xi_{i}+a\partial_{i}\dot{\xi}^S+\dot{a}\partial_{i}\xi^S)
\end{align}

\begin{align}
{\cal LS}_{C}&=- \frac{1}{a}(c_{1}+c_{3})\partial_{T}(a^{3}\bar{F}_K\partial_{(i}\xi_{j)})- (c_{1}+c_{3})\frac{1}{a}\partial_{T}(a^3\bar{F}_K\partial_{i} \partial_{j}\xi^S)+ a^{2}\delta_{ij}(-2+\alpha \bar{F}_K) H\dot{\phi}_P\nn\\
& - a^2\delta_{ij}\alpha H\bar{F}_K \partial_{k} \partial^{k}\xi^S-\frac{c_{2}}{a}\delta_{ij}\partial_{T}(a^{3}\bar{F}_K \partial_{k}\partial^{k}\xi^S)   \nn\\
&- 2a^{2}\delta_{ij} \phi_P(3H^{2}+\dot{H}(2-\alpha\bar{F}_K) - \alpha\frac{H}{a^{3}}\partial_{T}(a^{3}\bar{F}_K))- \delta_{ij}a^{2}(2-\alpha\bar{F}_K)\ddot{\psi}_P\nn\\
&+ 2aH \partial_{(i}\omega_{j)} +a\partial_{(i}\dot{\omega}_{j)}   -2a^{2}\delta_{ij}(8\pi G P + 3H^{2}+\dot{H}(2-\alpha\bar{F}_K)-\alpha\frac{H}{a^{3}}\partial_{T}(a^{3}\bar{F}_K))\psi_P\nn\\
&-a^{2}\delta_{ij}(6H - \alpha\frac{1}{a^{6}}\partial_{T}(a^{6}\bar{F}_K))\dot{\psi}_P 
\end{align}
\begin{align}
{\cal LS}_{D}&=-2c_{1}a\partial_{T}(\bar{F}_K(-\dot{\omega}_{i}+a\dot{\xi}_{i}+a\dot{\xi}_{i}+a\partial_{i}\dot{\xi}^S+\dot{a}\partial_{i}\xi^S)) \nn\\
&- 6c_{1}aH \bar{F}_K (a\partial_{i}\dot{\xi}^S+\dot{a}\partial_{i}\xi^S+a\dot{\xi}_{i}+\dot{a}\xi_{i}-\dot{\omega}_{i})\nn\\
& -a(2c_{1}-\alpha)\partial_{T}(H\bar{F}_K)(a\partial_{i}\xi^S+a\xi_{i}-\omega_{i}) -4c_{1}aH^{2}\bar{F}_K (a\partial_{i}\xi^S+a\xi_{i}-\omega_{i})  
\end{align}
\begin{align}
{\cal LS}_{A2}&=-c_{1}H \bar{F}_K\partial_{i}  \partial^{i}\xi^S +2\alpha H \bar{F}_K \partial_{i}\partial^{i}\xi^S  - \frac{c_{1}}{ a}\bar{F}_K \partial_{i} \partial^{i}\partial_T(a\xi^S) \nn\\
&+ (6H-6\alpha H\bar{F}_K)\dot{\psi}_P 
+16\pi G \rho\phi_P  -\frac{1}{2}\left(M^2F-M^2F(\bar{K})\right)\nn\\
&+3\alpha H^{2}\left(F_K-\bar{F}_K \right)-\phi_PM^2\bar{F}
\end{align}
\begin{align}
{\cal LS}_{B2}&= -a^2\bigg(2c_{1}H^{2}\bar{F}_K + (c_{1}-\alpha)\partial_{T}(H\bar{F}_K )\bigg)(\xi_{i}+\partial_{i}\xi^S)\nn\\
&-3c_{1}aH\bar{F}_K (-\dot{\omega}_{i}+a\dot{\xi}_{i}+\dot{a}\xi_{i}+a\partial_{i}\dot{\xi}^S+\dot{a}\partial_{i}\xi^S)-\frac{a\omega_i}{2}M^2\bar{F}\nn\\
&- c_{1}a\partial_{T}\bigg(\bar{F}_K(-\dot{\omega}_{i}+a\dot{\xi}_{i}+\dot{a}\xi_{i}+a \partial_{i}\dot{\xi}^S+\dot{a} \partial_{i} \xi^S)\bigg)\nn\\
&+\frac{1}{2}a\omega_{i}(-6H^{2}-4\dot{H}+16\pi G\rho+4c_{1}H^{2}\bar{F}_K+ 6\alpha H^{2}\bar{F}_K + 2c_{1}\partial_{T}(H\bar{F}_K))\label{eqn_vectorLS}\hspace{1cm}
\end{align}
\begin{align}
{\cal LS}_{C2}&= -\frac{1}{a}(c_{1}+c_{3})\partial_{T}(a^{3}\bar{F}_K\partial_{(i}\xi_{j)})-(c_{1}+c_{3})\frac{1}{a}\partial_{T}(a^3\bar{F}_K\partial_{i} \partial_{j}\xi^S)\nn\\
&+ a^{2}\delta_{ij}(-2+\alpha \bar{F}_K)H \dot{\phi}_P - a^2\delta_{ij}\alpha H\bar{F}_K \partial_{k} \partial^{k}\xi^S\nn\\
&-\frac{c_{2}}{a}\delta_{ij}\partial_{T}(a^{3}\bar{F}_K \partial_{k}\partial^{k}\xi^S)   - 2a^{2}\delta_{ij} \phi_P(3H^{2}+\dot{H}(2-\alpha\bar{F}_K) - \alpha\frac{H}{a^{3}}\partial_{T}(a^{3}\bar{F}_K))\nn\\
&- \delta_{ij}a^{2}(2-\alpha\bar{F}_K)\ddot{\psi}_P-2aH \partial_{(i}\omega_{j)} +a\partial_{(i}\dot{\omega}_{j)}  \nn\\
& - 2a^{2}\delta_{ij}(8\pi G P + 3H^{2}+\dot{H}(2-\alpha\bar{F}_K)+\alpha\frac{H}{a^{3}}\partial_{T}(a^{3}\bar{F}_K))\psi_P\nn\\
&-a^{2}\delta_{ij}(6H - \alpha\frac{1}{a^{6}}\partial_{T}(a^{6}\bar{F}_K))\dot{\psi}_P-a^2\psi_P\delta_{ij}M^2\bar{F} \nn\\
&+\delta_{ij}\bigg((F_K-\bar{F}_K)\left(-a\ddot{a}\alpha-2\dot{a}^2\alpha\right)+\frac{1}{2}\left(M^2F-M^2F(\bar{K})\right)- (\dot{F}_K-\dot{\bar{F}}_K)a\dot{a}\alpha \bigg)
\end{align}
\begin{align}
{\cal LS}_{D2}&=-2c_{1}a\partial_{T}(\bar{F}_K(-\dot{\omega}_{i}+a\dot{\xi}_{i}+a\dot{\xi}_{i}+a\partial_{i}\dot{\xi}^S+\dot{a}\partial_{i}\xi^S)) \nn\\
&- 6c_{1}aH \bar{F}_K (a\partial_{i}\dot{\xi}^S+\dot{a}\partial_{i}\xi^S+a\dot{\xi}_{i}+\dot{a}\xi_{i}-\dot{\omega}_{i})\nn\\
&-a(2c_{1}-\alpha)\partial_{T}(H\bar{F}_K)(a\partial_{i}\xi^S+a\xi_{i}-\omega_{i}) -4c_{1}aH^{2}\bar{F}_K (a\partial_{i}\xi^S+a\xi_{i}-\omega_{i})  
\end{align}

\section{Intermediate quantities in $\frac{1}{c}$ calculation}
\label{app_gory}
In this appendix we present some intermediate parts of our expansion in powers of $c$ for readers' reference, and to aid any attempts at checking, recreating or extending the derivation we have carried out.
\paragraph{Components of $K^{\alpha\beta}_{\ph{\alpha\beta}\mu\nu}$}
\begin{align}
K\indices{^0^0_0_0}&=c_1+c_2+c_3,\quad K\indices{^0^0_0_i}=ac_1\frac{B_i}{c^3},\quad K\indices{^0^0_i_0}=ac_1\frac{B_i}{c^3}\nonumber\\
K\indices{^0^0_i_j}&=a^2c_1\delta_{ij}\bigg(-1-2\frac{U_N}{c^2}-2\frac{V_N}{c^2} \bigg),\quad
K\indices{^0^k_0_0}=c_1\frac{B^k}{ac^3},\quad K\indices{^0^k_0_i}=c_2 \delta^k_i\nonumber\\
K\indices{^0^k_i_0}&=c_3 \delta^k_i,\quad K\indices{^0^k_i_j}=-c_1a\frac{B^k}{c^3}\delta_{ij},\quad K\indices{^k^0_0_0}=c_1\frac{B^k}{ac^3}\nonumber\\
K\indices{^k^0_0_i}&=c_3 \delta^k_i,\quad
K\indices{^k^0_i_0}=c_2 \delta^k_i,\quad
K\indices{^k^0_i_j}=-c_1a\frac{B^k}{c^3}\delta_{ij}\nonumber\\
K\indices{^l^k_0_0}&=\frac{c_1}{a^2}\delta^{lk}\bigg(-1+2\frac{U_N}{c^2} +2\frac{V_N}{c^2} \bigg),\quad
K\indices{^l^k_0_i}=-c_1\delta^{lk}\frac{B_i}{ac^3},\quad K\indices{^l^k_i_0}=-c_1\delta^{lk}\frac{B_i}{ac^3}\nonumber\\
K\indices{^l^k_i_j}&=c_1\delta^{lk}\delta_{ij}+c_2\delta^l_i\delta^k_j+c_3\delta^l_j\delta^k_i\nonumber
\end{align}
\paragraph{Components of $\nabla_{\mu}A^{\nu}$}
\begin{align}
\nabla_0 A^0 &=0,\quad \nabla_0 A^i=-\frac{U^{,i}_N}{a^2c^2},\quad
\nabla_i A^0=0\nn\\
\nabla_i A^j&=\delta_i^j\frac{\dot{a}}{ca}\left(1+\frac{U_N}{c^2} \right)+\frac{\beta^j_{,i}}{c^3}+\frac{B^{,j}_i-B^j_{,i}}{2ac^3}+\delta_{ij}\frac{\dot{V}_N}{c^3}\nonumber\\
\nabla_0 A_0 &=0,\quad \nabla_0 A_i=-\frac{U_{N,i}}{c^2},\quad\nabla_j A_0=0\nn\\
\nabla_i A_j&=\delta_{ij}\frac{a\dot{a}}{c}\left(1+\frac{U_N}{c^2}+2\frac{V_N}{c^2} \right)+\frac{a^2\beta_{j,i}}{c^3}+a\frac{B_{i,j}-B_{j,i}}{2c^3}+\delta_{ij}a^2\frac{\dot{V}_N}{c^3}\nonumber\\
\nabla^0 A_0&=0,\quad \nabla^0 A_i=\frac{U_{N,i}}{c^2},\quad \nabla^i A_0=0\nonumber\\
\nabla^i A_j&=\delta^i_{j}\frac{\dot{a}}{ca}\left(1+\frac{U_N}{c^2} \right)+\frac{\beta^{,i}_{j}}{c^3}+\frac{B^i_{,j}-B^{,i}_{j}}{2ac^3}+\delta_{ij}\frac{\dot{V}_N}{c^3}
\nonumber
\end{align}
\paragraph{Components of $J^{\mu}_{\ph{\mu}\nu}$}
\begin{align}
J\indices{^0_0}&=6\frac{c_2\dot{a}}{ca}+6\frac{c_2\dot{a}U_N}{c^3a}+2\frac{c_2}{c^3}\beta^i_{,i}+\frac{6c_2\dot{V}_N}{c^3},\quad J\indices{^0_i}=2\frac{c_1 U_{N,i}}{c^2},\quad J\indices{^i_0}=-2\frac{c_3 U_{N,i}}{a^2c^2}\nonumber\\
J\indices{^i_j}&=2\delta^i_j\frac{\alpha\dot{a}}{ca}+2\delta^i_j\frac{\alpha\dot{a}U_N}{c^3a}+2\delta^i_j\frac{c_2\beta^k_{,k}}{c^3}+2\frac{c_1\beta^{,i}_{j}}{c^3}+2\frac{c_3\beta^{i}_{,j}}{c^3}+\frac{2\alpha\delta^i_j\dot{V}_N}{c^3}+\frac{1}{ac^3}\left(c_1-c_3\right)\left(B^i_{,j}-B^{,i}_{j} \right)\nonumber\\
J_{00}&=-6\frac{c_2\dot{a}}{ca}+6\frac{c_2\dot{a}U_N}{c^3a}-2\frac{c_2}{c^3}\beta^k_{,k}-\frac{6c_2\dot{V}_N}{c^3},\quad J_{0i}=-2\frac{c_1 U_{N,i}}{c^2},\quad J_{i0}=-2\frac{c_3 U_{N,i}}{c^2}\nonumber\\
J_{ij}&=2\delta_{ij}\frac{\alpha a\dot{a}}{c}+2\delta_{ij}\frac{\alpha a\dot{a}U_N}{c^3}+4\delta_{ij}\frac{\alpha a\dot{a}V_N}{c^3}+2\delta_{ij}a^2\frac{c_2\beta^k_{,k}}{c^3}+2a^2c_1\frac{\beta_{j,i}}{c^3}+\frac{2\alpha a^2 \delta_{ij}\dot{V}_N}{c^3}\nn\\
&+2a^2c_3\frac{\beta_{i,j}}{c^3}+\frac{a}{c^3}\left(c_1-c_3\right)\left(B_{i,j}-B_{j,i} \right)\nonumber\\
J\indices{_0^0}&=6\frac{c_2\dot{a}}{ca}+6\frac{c_2\dot{a}U_N}{c^3a}+2\frac{c_2}{c^3}\beta^i_{,i}+\frac{6c_2\dot{V}_N}{c^3},\quad J\indices{_0^i}=-2\frac{c_1 U_{N,i}}{a^2c^2},\quad J\indices{_i^0}=2\frac{c_3 U_{N,i}}{c^2}\nonumber\\
J\indices{_i^j}&=2\delta_{i}^j\frac{\alpha \dot{a}}{ac}+2\delta_{i}^j\frac{\alpha \dot{a}U_N}{ac^3}+2\delta_{i}^j\frac{c_2\beta^k_{,k}}{c^3}+2c_1\frac{\beta_{,i}^{j}}{c^3}+2c_3\frac{\beta_{i}^{,j}}{c^3}+\frac{2\alpha \delta_{ij}\dot{V}_N}{c^3}+\frac{1}{ac^3}\left(c_1-c_3\right)\left(B^{,j}_{i}-B^j_{,i} \right)\nonumber
\end{align}
\paragraph{The scalars $K$ and $\lambda$ and the tensor $Y_{\mu\nu}$}

\begin{align}
K&=3\frac{\alpha\dot{a}^2}{M^2c^2a^2}+2\frac{\alpha\dot{a}\beta^i_{,i}}{M^2ac^4}-c_1\frac{U_{N,k}U^{,k}_N}{M^2a^2c^4}+6\frac{\alpha \dot{a}^2U_{N}}{M^2a^2c^4}+\frac{6\dot{a}\alpha\dot{V}_N}{M^2ac^4}\\
\lambda&=-3\frac{c_2 \dot{a}}{c^2a}\dot{F}_K+\frac{c_3F_{K,i}U_{N,i}}{a^2c^2}+\frac{F_k}{c^2}\bigg(\frac{c_3 U_{N,kk}}{a^2}+3\frac{\alpha\dot{a}^2}{a^2}-6\frac{c_2 \dot{a}^2}{a^2}-3\frac{c_2\ddot{a}}{a} \bigg) \\
Y_{00}&=0,\quad Y_{0i}=-\frac{c_1\dot{a}U_{N,i}}{ac^3},\quad Y_{i0}=-\frac{c_1\dot{a}U_{N,i}}{ac^3},\quad Y_{ij}=0
\end{align}

\paragraph{The tensor $\nabla_{\mu}J_{\nu}^{\ph{\nu}\alpha}$}
\begin{align}
\nabla_\sigma J\indices{_0^\sigma}&=12\frac{c_2}{c^2}H^{2}+6\frac{c_2}{c^2}\frac{\ddot{a}}{a}-6\frac{\alpha}{c^2}H^{2}-2\frac{c_1}{a^2c^2}U^{,i}_{N,i}\nonumber\\
\nabla_\sigma J\indices{_i^\sigma}&=2\frac{c_3\dot{U}_{N,i}}{c^3}+2(\alpha+c_3)\frac{U_{N,i}\dot{a}}{c^3a}+\frac{2\alpha\dot{V}_{N,i}}{c^3}\nonumber\\
&+\partial_k\bigg(2\delta^k_i \frac{c_2 \beta^l_{,l}}{c^3}+2\frac{c_1\beta^{k}_{,i}}{c^3}+2\frac{c_3\beta^{,k}_i}{c^3}+\frac{(c_1-c_3)}{ac^3}\left(B^{,k}_i-B^k_{,i} \right) \bigg)\nonumber\\
\nabla_\sigma J\indices{^\sigma_0}&=12\frac{c_2}{c^2}H^{2}+6\frac{c_2}{c^2}\frac{\ddot{a}}{a}-6\frac{\alpha}{c^2}H^{2}-2\frac{c_3}{a^2c^2}U^{,i}_{N,i}\nonumber\\
\nabla_\sigma J\indices{^\sigma_j}&=2\frac{c_1\dot{U}_{N,j}}{c^3}+2(\alpha+c_1)\frac{U_{N,j}\dot{a}}{c^3a}+\frac{2\alpha\dot{V}_{N,i}}{c^3}\nonumber\\
&+\partial_i\bigg(2\delta^i_j\frac{c_2\beta^k_{,k}}{c^3}+2\frac{c_1\beta^{,i}_{j}}{c^3}+2\frac{c_3\beta^{i}_{,j}}{c^3}+\frac{1}{ac^3}\left(c_1-c_3\right)\left(B^i_{,j}-B^{,i}_{j} \right) \bigg)\nonumber\\
\nabla_0 J\indices{_0_0}&=6\frac{c_2}{c^2}\bigg(\frac{\ddot{a}}{a}- H^{2}\bigg)\nonumber\\
\nabla_i J\indices{_0_0}&=-6\frac{c_2 \dot{a}U_{N,i}}{ac^3}-2\frac{c_2}{c^3}\beta^k_{,ki}+2(c_1+c_3)\frac{U_{N,i}\dot{a}}{c^3a}-\frac{6c_2\dot{V}_{N,i}}{c^3}\nonumber\\
\nabla_0 J\indices{_0_i}&=-2\frac{c_1\dot{U}_{N,i}}{c^3}+2\left(2c_1+c_3 \right)\frac{U_{N,i}\dot{a}}{c^3a}\nonumber\\
\nabla_0 J\indices{_i_0}&=-2\frac{c_3\dot{U}_{N,i}}{c^3}+2\left(c_1+2c_3 \right)\frac{U_{N,i}\dot{a}}{c^3a}\nonumber\\
\nabla_j J\indices{_0_i}&=-2\frac{c_1 U_{N,ij}}{c^2}+\delta_{ij}\frac{\dot{a}^2}{c^2}\left(6c_2-2\alpha\right)\nonumber\\
\nabla_j J\indices{_i_0}&=-2\frac{c_3 U_{N,ij}}{c^2}+\delta_{ij}\frac{\dot{a}^2}{c^2}\left(6c_2-2\alpha\right)\nonumber\\
\nabla_0 J\indices{_i_j}&=2\delta_{ij}\frac{\alpha}{c^2}\left(a\ddot{a}-\dot{a}^2 \right)\nonumber \\
\nabla_i J\indices{_j_k}&=\frac{2c_1 a \dot{a}\delta_{ij}U_{N,k}}{c^3}+\frac{2c_3 a \dot{a}\delta_{ik}U_{N,j}}{c^3}+\frac{2\alpha a \dot{a} \delta_{jk}U_{N,i}}{c^3}+\frac{2a^2c_2 \delta_{jk}\beta^l_{,li}}{c^3}\nonumber\\
& +\frac{2\alpha a^2 \delta_{ik} \dot{V}_{N,i}}{c^3} +\frac{2a^2c_1\beta_{k,ij}}{c^3}+\frac{2a^2c_3\beta_{j,ik}}{c^3}+\frac{a\left(c_1-c_3 \right)}{c^3}\left(B_{j,ki}-B_{k,ji} \right) \nonumber
\end{align}

\bibliography{postfmond}

\begin{thebibliography}{66}%
\makeatletter
\providecommand \@ifxundefined [1]{%
 \@ifx{#1\undefined}
}%
\providecommand \@ifnum [1]{%
 \ifnum #1\expandafter \@firstoftwo
 \else \expandafter \@secondoftwo
 \fi
}%
\providecommand \@ifx [1]{%
 \ifx #1\expandafter \@firstoftwo
 \else \expandafter \@secondoftwo
 \fi
}%
\providecommand \natexlab [1]{#1}%
\providecommand \enquote  [1]{``#1''}%
\providecommand \bibnamefont  [1]{#1}%
\providecommand \bibfnamefont [1]{#1}%
\providecommand \citenamefont [1]{#1}%
\providecommand \href@noop [0]{\@secondoftwo}%
\providecommand \href [0]{\begingroup \@sanitize@url \@href}%
\providecommand \@href[1]{\@@startlink{#1}\@@href}%
\providecommand \@@href[1]{\endgroup#1\@@endlink}%
\providecommand \@sanitize@url [0]{\catcode `\\12\catcode `\$12\catcode
  `\&12\catcode `\#12\catcode `\^12\catcode `\_12\catcode `\%12\relax}%
\providecommand \@@startlink[1]{}%
\providecommand \@@endlink[0]{}%
\providecommand \url  [0]{\begingroup\@sanitize@url \@url }%
\providecommand \@url [1]{\endgroup\@href {#1}{\urlprefix }}%
\providecommand \urlprefix  [0]{URL }%
\providecommand \Eprint [0]{\href }%
\providecommand \doibase [0]{http://dx.doi.org/}%
\providecommand \selectlanguage [0]{\@gobble}%
\providecommand \bibinfo  [0]{\@secondoftwo}%
\providecommand \bibfield  [0]{\@secondoftwo}%
\providecommand \translation [1]{[#1]}%
\providecommand \BibitemOpen [0]{}%
\providecommand \bibitemStop [0]{}%
\providecommand \bibitemNoStop [0]{.\EOS\space}%
\providecommand \EOS [0]{\spacefactor3000\relax}%
\providecommand \BibitemShut  [1]{\csname bibitem#1\endcsname}%
\let\auto@bib@innerbib\@empty
\bibitem [{\citenamefont {Aghanim}\ \emph {et~al.}(2020)\citenamefont {Aghanim}
  \emph {et~al.}}]{planck}%
  \BibitemOpen
  \bibfield  {author} {\bibinfo {author} {\bibfnamefont {N.}~\bibnamefont
  {Aghanim}} \emph {et~al.} (\bibinfo {collaboration} {Planck}),\ }\href
  {\doibase 10.1051/0004-6361/201833910} {\bibfield  {journal} {\bibinfo
  {journal} {Astron. Astrophys.}\ }\textbf {\bibinfo {volume} {641}},\ \bibinfo
  {pages} {A6} (\bibinfo {year} {2020})},\ \bibinfo {note} {[Erratum:
  Astron.Astrophys. 652, C4 (2021)]},\ \Eprint
  {http://arxiv.org/abs/1807.06209} {arXiv:1807.06209 [astro-ph.CO]}
  \BibitemShut {NoStop}%
\bibitem [{\citenamefont {Del~Popolo}\ and\ \citenamefont
  {Le~Delliou}(2017)}]{1606.07790}%
  \BibitemOpen
  \bibfield  {author} {\bibinfo {author} {\bibfnamefont {A.}~\bibnamefont
  {Del~Popolo}}\ and\ \bibinfo {author} {\bibfnamefont {M.}~\bibnamefont
  {Le~Delliou}},\ }\href {\doibase 10.3390/galaxies5010017} {\bibfield
  {journal} {\bibinfo  {journal} {Galaxies}\ }\textbf {\bibinfo {volume} {5}},\
  \bibinfo {pages} {17} (\bibinfo {year} {2017})},\ \Eprint
  {http://arxiv.org/abs/1606.07790} {arXiv:1606.07790 [astro-ph.CO]}
  \BibitemShut {NoStop}%
\bibitem [{\citenamefont {Bullock}\ and\ \citenamefont
  {Boylan-Kolchin}(2017)}]{1707.04256}%
  \BibitemOpen
  \bibfield  {author} {\bibinfo {author} {\bibfnamefont {J.~S.}\ \bibnamefont
  {Bullock}}\ and\ \bibinfo {author} {\bibfnamefont {M.}~\bibnamefont
  {Boylan-Kolchin}},\ }\href {\doibase 10.1146/annurev-astro-091916-055313}
  {\bibfield  {journal} {\bibinfo  {journal} {Ann. Rev. Astron. Astrophys.}\
  }\textbf {\bibinfo {volume} {55}},\ \bibinfo {pages} {343} (\bibinfo {year}
  {2017})},\ \Eprint {http://arxiv.org/abs/1707.04256} {arXiv:1707.04256
  [astro-ph.CO]} \BibitemShut {NoStop}%
\bibitem [{\citenamefont {Di~Valentino}\ \emph {et~al.}(2021)\citenamefont
  {Di~Valentino}, \citenamefont {Mena}, \citenamefont {Pan}, \citenamefont
  {Visinelli}, \citenamefont {Yang}, \citenamefont {Melchiorri}, \citenamefont
  {Mota}, \citenamefont {Riess},\ and\ \citenamefont {Silk}}]{2103.01183}%
  \BibitemOpen
  \bibfield  {author} {\bibinfo {author} {\bibfnamefont {E.}~\bibnamefont
  {Di~Valentino}}, \bibinfo {author} {\bibfnamefont {O.}~\bibnamefont {Mena}},
  \bibinfo {author} {\bibfnamefont {S.}~\bibnamefont {Pan}}, \bibinfo {author}
  {\bibfnamefont {L.}~\bibnamefont {Visinelli}}, \bibinfo {author}
  {\bibfnamefont {W.}~\bibnamefont {Yang}}, \bibinfo {author} {\bibfnamefont
  {A.}~\bibnamefont {Melchiorri}}, \bibinfo {author} {\bibfnamefont {D.~F.}\
  \bibnamefont {Mota}}, \bibinfo {author} {\bibfnamefont {A.~G.}\ \bibnamefont
  {Riess}}, \ and\ \bibinfo {author} {\bibfnamefont {J.}~\bibnamefont {Silk}},\
  }\href {\doibase 10.1088/1361-6382/ac086d} {\bibfield  {journal} {\bibinfo
  {journal} {Class. Quant. Grav.}\ }\textbf {\bibinfo {volume} {38}},\ \bibinfo
  {pages} {153001} (\bibinfo {year} {2021})},\ \Eprint
  {http://arxiv.org/abs/2103.01183} {arXiv:2103.01183 [astro-ph.CO]}
  \BibitemShut {NoStop}%
\bibitem [{\citenamefont {Perivolaropoulos}\ and\ \citenamefont
  {Skara}(2022)}]{2105.05208}%
  \BibitemOpen
  \bibfield  {author} {\bibinfo {author} {\bibfnamefont {L.}~\bibnamefont
  {Perivolaropoulos}}\ and\ \bibinfo {author} {\bibfnamefont {F.}~\bibnamefont
  {Skara}},\ }\href {\doibase 10.1016/j.newar.2022.101659} {\bibfield
  {journal} {\bibinfo  {journal} {New Astron. Rev.}\ }\textbf {\bibinfo
  {volume} {95}},\ \bibinfo {pages} {101659} (\bibinfo {year} {2022})},\
  \Eprint {http://arxiv.org/abs/2105.05208} {arXiv:2105.05208 [astro-ph.CO]}
  \BibitemShut {NoStop}%
\bibitem [{\citenamefont {{Clifton}}\ \emph {et~al.}(2012)\citenamefont
  {{Clifton}}, \citenamefont {{Ferreira}}, \citenamefont {{Padilla}},\ and\
  \citenamefont {{Skordis}}}]{1106.2476}%
  \BibitemOpen
  \bibfield  {author} {\bibinfo {author} {\bibfnamefont {T.}~\bibnamefont
  {{Clifton}}}, \bibinfo {author} {\bibfnamefont {P.~G.}\ \bibnamefont
  {{Ferreira}}}, \bibinfo {author} {\bibfnamefont {A.}~\bibnamefont
  {{Padilla}}}, \ and\ \bibinfo {author} {\bibfnamefont {C.}~\bibnamefont
  {{Skordis}}},\ }\href {\doibase 10.1016/j.physrep.2012.01.001} {\bibfield
  {journal} {\bibinfo  {journal} {\physrep}\ }\textbf {\bibinfo {volume}
  {513}},\ \bibinfo {pages} {1} (\bibinfo {year} {2012})},\ \Eprint
  {http://arxiv.org/abs/1106.2476} {arXiv:1106.2476} \BibitemShut {NoStop}%
\bibitem [{\citenamefont {{Milgrom}}(1983{\natexlab{a}})}]{milgrom1}%
  \BibitemOpen
  \bibfield  {author} {\bibinfo {author} {\bibfnamefont {M.}~\bibnamefont
  {{Milgrom}}},\ }\href {\doibase 10.1086/161130} {\bibfield  {journal}
  {\bibinfo  {journal} {\apj}\ }\textbf {\bibinfo {volume} {270}},\ \bibinfo
  {pages} {365} (\bibinfo {year} {1983}{\natexlab{a}})}\BibitemShut {NoStop}%
\bibitem [{\citenamefont {{Milgrom}}(1983{\natexlab{b}})}]{milgrom2}%
  \BibitemOpen
  \bibfield  {author} {\bibinfo {author} {\bibfnamefont {M.}~\bibnamefont
  {{Milgrom}}},\ }\href {\doibase 10.1086/161131} {\bibfield  {journal}
  {\bibinfo  {journal} {\apj}\ }\textbf {\bibinfo {volume} {270}},\ \bibinfo
  {pages} {371} (\bibinfo {year} {1983}{\natexlab{b}})}\BibitemShut {NoStop}%
\bibitem [{\citenamefont {{Milgrom}}(1983{\natexlab{c}})}]{milgrom3}%
  \BibitemOpen
  \bibfield  {author} {\bibinfo {author} {\bibfnamefont {M.}~\bibnamefont
  {{Milgrom}}},\ }\href {\doibase 10.1086/161132} {\bibfield  {journal}
  {\bibinfo  {journal} {\apj}\ }\textbf {\bibinfo {volume} {270}},\ \bibinfo
  {pages} {384} (\bibinfo {year} {1983}{\natexlab{c}})}\BibitemShut {NoStop}%
\bibitem [{\citenamefont {Milgrom}(2001)}]{milgromreview}%
  \BibitemOpen
  \bibfield  {author} {\bibinfo {author} {\bibfnamefont {M.}~\bibnamefont
  {Milgrom}},\ }\href@noop {} {\bibfield  {journal} {\bibinfo  {journal} {Acta
  Phys. Polon. B}\ }\textbf {\bibinfo {volume} {32}},\ \bibinfo {pages} {3613}
  (\bibinfo {year} {2001})},\ \Eprint {http://arxiv.org/abs/astro-ph/0112069}
  {arXiv:astro-ph/0112069} \BibitemShut {NoStop}%
\bibitem [{\citenamefont {Famaey}\ and\ \citenamefont
  {McGaugh}(2012)}]{famaeyreview}%
  \BibitemOpen
  \bibfield  {author} {\bibinfo {author} {\bibfnamefont {B.}~\bibnamefont
  {Famaey}}\ and\ \bibinfo {author} {\bibfnamefont {S.}~\bibnamefont
  {McGaugh}},\ }\href {\doibase 10.12942/lrr-2012-10} {\bibfield  {journal}
  {\bibinfo  {journal} {Living Rev. Rel.}\ }\textbf {\bibinfo {volume} {15}},\
  \bibinfo {pages} {10} (\bibinfo {year} {2012})},\ \Eprint
  {http://arxiv.org/abs/1112.3960} {arXiv:1112.3960 [astro-ph.CO]} \BibitemShut
  {NoStop}%
\bibitem [{\citenamefont {Diaferio}\ and\ \citenamefont
  {Angus}(2016)}]{1206.6231}%
  \BibitemOpen
  \bibfield  {author} {\bibinfo {author} {\bibfnamefont {A.}~\bibnamefont
  {Diaferio}}\ and\ \bibinfo {author} {\bibfnamefont {G.~W.}\ \bibnamefont
  {Angus}},\ }\enquote {\bibinfo {title} {{The Acceleration Scale, Modified
  Newtonian Dynamics, and Sterile Neutrinos}},}\ in\ \href {\doibase
  10.1007/978-3-319-20224-2_10} {\emph {\bibinfo {booktitle} {{Gravity: Where
  Do We Stand?}}}},\ \bibinfo {editor} {edited by\ \bibinfo {editor}
  {\bibfnamefont {R.}~\bibnamefont {Peron}}, \bibinfo {editor} {\bibfnamefont
  {M.}~\bibnamefont {Colpi}}, \bibinfo {editor} {\bibfnamefont
  {V.}~\bibnamefont {Gorini}}, \ and\ \bibinfo {editor} {\bibfnamefont
  {U.}~\bibnamefont {Moschella}}}\ (\bibinfo {year} {2016})\ pp.\ \bibinfo
  {pages} {337--366},\ \Eprint {http://arxiv.org/abs/1206.6231}
  {arXiv:1206.6231 [astro-ph.CO]} \BibitemShut {NoStop}%
\bibitem [{\citenamefont {Banik}\ and\ \citenamefont
  {Zhao}(2022)}]{banikreview}%
  \BibitemOpen
  \bibfield  {author} {\bibinfo {author} {\bibfnamefont {I.}~\bibnamefont
  {Banik}}\ and\ \bibinfo {author} {\bibfnamefont {H.}~\bibnamefont {Zhao}},\
  }\href {\doibase 10.3390/sym14071331} {\bibfield  {journal} {\bibinfo
  {journal} {Symmetry}\ }\textbf {\bibinfo {volume} {14}},\ \bibinfo {pages}
  {1331} (\bibinfo {year} {2022})},\ \Eprint {http://arxiv.org/abs/2110.06936}
  {arXiv:2110.06936 [astro-ph.CO]} \BibitemShut {NoStop}%
\bibitem [{\citenamefont {Hees}\ \emph {et~al.}(2016)\citenamefont {Hees},
  \citenamefont {Famaey}, \citenamefont {Angus},\ and\ \citenamefont
  {Gentile}}]{1510.01369}%
  \BibitemOpen
  \bibfield  {author} {\bibinfo {author} {\bibfnamefont {A.}~\bibnamefont
  {Hees}}, \bibinfo {author} {\bibfnamefont {B.}~\bibnamefont {Famaey}},
  \bibinfo {author} {\bibfnamefont {G.~W.}\ \bibnamefont {Angus}}, \ and\
  \bibinfo {author} {\bibfnamefont {G.}~\bibnamefont {Gentile}},\ }\href
  {\doibase 10.1093/mnras/stv2330} {\bibfield  {journal} {\bibinfo  {journal}
  {Mon. Not. Roy. Astron. Soc.}\ }\textbf {\bibinfo {volume} {455}},\ \bibinfo
  {pages} {449} (\bibinfo {year} {2016})},\ \Eprint
  {http://arxiv.org/abs/1510.01369} {arXiv:1510.01369 [astro-ph.GA]}
  \BibitemShut {NoStop}%
\bibitem [{\citenamefont {{McGaugh}}\ \emph {et~al.}(2016)\citenamefont
  {{McGaugh}}, \citenamefont {{Lelli}},\ and\ \citenamefont
  {{Schombert}}}]{1609.05917}%
  \BibitemOpen
  \bibfield  {author} {\bibinfo {author} {\bibfnamefont {S.~S.}\ \bibnamefont
  {{McGaugh}}}, \bibinfo {author} {\bibfnamefont {F.}~\bibnamefont {{Lelli}}},
  \ and\ \bibinfo {author} {\bibfnamefont {J.~M.}\ \bibnamefont
  {{Schombert}}},\ }\href {\doibase 10.1103/PhysRevLett.117.201101} {\bibfield
  {journal} {\bibinfo  {journal} {\prl}\ }\textbf {\bibinfo {volume} {117}},\
  \bibinfo {eid} {201101} (\bibinfo {year} {2016})},\ \Eprint
  {http://arxiv.org/abs/1609.05917} {arXiv:1609.05917 [astro-ph.GA]}
  \BibitemShut {NoStop}%
\bibitem [{\citenamefont {{Sanders}}(2019)}]{1811.05260}%
  \BibitemOpen
  \bibfield  {author} {\bibinfo {author} {\bibfnamefont {R.~H.}\ \bibnamefont
  {{Sanders}}},\ }\href {\doibase 10.1093/mnras/stz353} {\bibfield  {journal}
  {\bibinfo  {journal} {\mnras}\ }\textbf {\bibinfo {volume} {485}},\ \bibinfo
  {pages} {513} (\bibinfo {year} {2019})},\ \Eprint
  {http://arxiv.org/abs/1811.05260} {arXiv:1811.05260 [astro-ph.GA]}
  \BibitemShut {NoStop}%
\bibitem [{\citenamefont {Lelli}\ \emph {et~al.}(2019)\citenamefont {Lelli},
  \citenamefont {McGaugh}, \citenamefont {Schombert}, \citenamefont {Desmond},\
  and\ \citenamefont {Katz}}]{1901.05966}%
  \BibitemOpen
  \bibfield  {author} {\bibinfo {author} {\bibfnamefont {F.}~\bibnamefont
  {Lelli}}, \bibinfo {author} {\bibfnamefont {S.~S.}\ \bibnamefont {McGaugh}},
  \bibinfo {author} {\bibfnamefont {J.~M.}\ \bibnamefont {Schombert}}, \bibinfo
  {author} {\bibfnamefont {H.}~\bibnamefont {Desmond}}, \ and\ \bibinfo
  {author} {\bibfnamefont {H.}~\bibnamefont {Katz}},\ }\href {\doibase
  10.1093/mnras/stz205} {\bibfield  {journal} {\bibinfo  {journal} {Mon. Not.
  Roy. Astron. Soc.}\ }\textbf {\bibinfo {volume} {484}},\ \bibinfo {pages}
  {3267} (\bibinfo {year} {2019})},\ \Eprint {http://arxiv.org/abs/1901.05966}
  {arXiv:1901.05966 [astro-ph.GA]} \BibitemShut {NoStop}%
\bibitem [{\citenamefont {Chae}\ \emph {et~al.}(2020)\citenamefont {Chae},
  \citenamefont {Lelli}, \citenamefont {Desmond}, \citenamefont {McGaugh},
  \citenamefont {Li},\ and\ \citenamefont {Schombert}}]{2009.11525}%
  \BibitemOpen
  \bibfield  {author} {\bibinfo {author} {\bibfnamefont {K.-H.}\ \bibnamefont
  {Chae}}, \bibinfo {author} {\bibfnamefont {F.}~\bibnamefont {Lelli}},
  \bibinfo {author} {\bibfnamefont {H.}~\bibnamefont {Desmond}}, \bibinfo
  {author} {\bibfnamefont {S.~S.}\ \bibnamefont {McGaugh}}, \bibinfo {author}
  {\bibfnamefont {P.}~\bibnamefont {Li}}, \ and\ \bibinfo {author}
  {\bibfnamefont {J.~M.}\ \bibnamefont {Schombert}},\ }\href {\doibase
  10.3847/1538-4357/abbb96} {\bibfield  {journal} {\bibinfo  {journal}
  {Astrophys. J.}\ }\textbf {\bibinfo {volume} {904}},\ \bibinfo {pages} {51}
  (\bibinfo {year} {2020})},\ \bibinfo {note} {[Erratum: Astrophys.J. 910, 81
  (2021)]},\ \Eprint {http://arxiv.org/abs/2009.11525} {arXiv:2009.11525
  [astro-ph.GA]} \BibitemShut {NoStop}%
\bibitem [{\citenamefont {Chae}\ \emph {et~al.}(2021)\citenamefont {Chae},
  \citenamefont {Desmond}, \citenamefont {Lelli}, \citenamefont {McGaugh},\
  and\ \citenamefont {Schombert}}]{2109.04745}%
  \BibitemOpen
  \bibfield  {author} {\bibinfo {author} {\bibfnamefont {K.-H.}\ \bibnamefont
  {Chae}}, \bibinfo {author} {\bibfnamefont {H.}~\bibnamefont {Desmond}},
  \bibinfo {author} {\bibfnamefont {F.}~\bibnamefont {Lelli}}, \bibinfo
  {author} {\bibfnamefont {S.~S.}\ \bibnamefont {McGaugh}}, \ and\ \bibinfo
  {author} {\bibfnamefont {J.~M.}\ \bibnamefont {Schombert}},\ }\href {\doibase
  10.3847/1538-4357/ac1bba} {\bibfield  {journal} {\bibinfo  {journal}
  {Astrophys. J.}\ }\textbf {\bibinfo {volume} {921}},\ \bibinfo {pages} {104}
  (\bibinfo {year} {2021})},\ \Eprint {http://arxiv.org/abs/2109.04745}
  {arXiv:2109.04745 [astro-ph.GA]} \BibitemShut {NoStop}%
\bibitem [{\citenamefont {Paranjape}\ and\ \citenamefont
  {Sheth}(2022)}]{2112.00026}%
  \BibitemOpen
  \bibfield  {author} {\bibinfo {author} {\bibfnamefont {A.}~\bibnamefont
  {Paranjape}}\ and\ \bibinfo {author} {\bibfnamefont {R.~K.}\ \bibnamefont
  {Sheth}},\ }\href {\doibase 10.1093/mnras/stac2689} {\bibfield  {journal}
  {\bibinfo  {journal} {Mon. Not. Roy. Astron. Soc.}\ }\textbf {\bibinfo
  {volume} {517}},\ \bibinfo {pages} {130} (\bibinfo {year} {2022})},\ \Eprint
  {http://arxiv.org/abs/2112.00026} {arXiv:2112.00026 [astro-ph.CO]}
  \BibitemShut {NoStop}%
\bibitem [{\citenamefont {Lelli}(2022)}]{2201.11752}%
  \BibitemOpen
  \bibfield  {author} {\bibinfo {author} {\bibfnamefont {F.}~\bibnamefont
  {Lelli}},\ }\href {\doibase 10.1038/s41550-021-01562-2} {\bibfield  {journal}
  {\bibinfo  {journal} {Nature Astron.}\ }\textbf {\bibinfo {volume} {6}},\
  \bibinfo {pages} {35} (\bibinfo {year} {2022})},\ \Eprint
  {http://arxiv.org/abs/2201.11752} {arXiv:2201.11752 [astro-ph.GA]}
  \BibitemShut {NoStop}%
\bibitem [{\citenamefont {Haslbauer}\ \emph {et~al.}(2022)\citenamefont
  {Haslbauer}, \citenamefont {Banik}, \citenamefont {Kroupa}, \citenamefont
  {Wittenburg},\ and\ \citenamefont {Javanmardi}}]{2202.01221}%
  \BibitemOpen
  \bibfield  {author} {\bibinfo {author} {\bibfnamefont {M.}~\bibnamefont
  {Haslbauer}}, \bibinfo {author} {\bibfnamefont {I.}~\bibnamefont {Banik}},
  \bibinfo {author} {\bibfnamefont {P.}~\bibnamefont {Kroupa}}, \bibinfo
  {author} {\bibfnamefont {N.}~\bibnamefont {Wittenburg}}, \ and\ \bibinfo
  {author} {\bibfnamefont {B.}~\bibnamefont {Javanmardi}},\ }\href {\doibase
  10.3847/1538-4357/ac46ac} {\bibfield  {journal} {\bibinfo  {journal}
  {Astrophys. J.}\ }\textbf {\bibinfo {volume} {925}},\ \bibinfo {pages} {183}
  (\bibinfo {year} {2022})},\ \Eprint {http://arxiv.org/abs/2202.01221}
  {arXiv:2202.01221 [astro-ph.GA]} \BibitemShut {NoStop}%
\bibitem [{\citenamefont {Lopez-Corredoira}\ \emph {et~al.}(2022)\citenamefont
  {Lopez-Corredoira}, \citenamefont {Betancort-Rijo}, \citenamefont {Scarpa},\
  and\ \citenamefont {Chrobakova}}]{2210.13961}%
  \BibitemOpen
  \bibfield  {author} {\bibinfo {author} {\bibfnamefont {M.}~\bibnamefont
  {Lopez-Corredoira}}, \bibinfo {author} {\bibfnamefont {J.~E.}\ \bibnamefont
  {Betancort-Rijo}}, \bibinfo {author} {\bibfnamefont {R.}~\bibnamefont
  {Scarpa}}, \ and\ \bibinfo {author} {\bibfnamefont {Z.}~\bibnamefont
  {Chrobakova}},\ }\href {\doibase 10.1093/mnras/stac3117} {\  (\bibinfo {year}
  {2022}),\ 10.1093/mnras/stac3117},\ \Eprint {http://arxiv.org/abs/2210.13961}
  {arXiv:2210.13961 [astro-ph.GA]} \BibitemShut {NoStop}%
\bibitem [{\citenamefont {{Bekenstein}}\ and\ \citenamefont
  {{Milgrom}}(1984)}]{bekensteinmilgrom1984}%
  \BibitemOpen
  \bibfield  {author} {\bibinfo {author} {\bibfnamefont {J.}~\bibnamefont
  {{Bekenstein}}}\ and\ \bibinfo {author} {\bibfnamefont {M.}~\bibnamefont
  {{Milgrom}}},\ }\href {\doibase 10.1086/162570} {\bibfield  {journal}
  {\bibinfo  {journal} {\apj}\ }\textbf {\bibinfo {volume} {286}},\ \bibinfo
  {pages} {7} (\bibinfo {year} {1984})}\BibitemShut {NoStop}%
\bibitem [{\citenamefont {Bekenstein}(2004)}]{0403694}%
  \BibitemOpen
  \bibfield  {author} {\bibinfo {author} {\bibfnamefont {J.~D.}\ \bibnamefont
  {Bekenstein}},\ }\href {\doibase 10.1103/PhysRevD.70.083509} {\bibfield
  {journal} {\bibinfo  {journal} {Phys. Rev. D}\ }\textbf {\bibinfo {volume}
  {70}},\ \bibinfo {pages} {083509} (\bibinfo {year} {2004})},\ \bibinfo {note}
  {[Erratum: Phys.Rev.D 71, 069901 (2005)]},\ \Eprint
  {http://arxiv.org/abs/astro-ph/0403694} {arXiv:astro-ph/0403694} \BibitemShut
  {NoStop}%
\bibitem [{\citenamefont {Bekenstein}(2011)}]{1201.2759}%
  \BibitemOpen
  \bibfield  {author} {\bibinfo {author} {\bibfnamefont {J.~D.}\ \bibnamefont
  {Bekenstein}},\ }\href {\doibase 10.1098/rsta.2011.0282} {\bibfield
  {journal} {\bibinfo  {journal} {Phil. Trans. Roy. Soc. Lond. A}\ }\textbf
  {\bibinfo {volume} {369}},\ \bibinfo {pages} {5003} (\bibinfo {year}
  {2011})},\ \Eprint {http://arxiv.org/abs/1201.2759} {arXiv:1201.2759
  [astro-ph.CO]} \BibitemShut {NoStop}%
\bibitem [{\citenamefont {Milgrom}(2009)}]{0912.0790}%
  \BibitemOpen
  \bibfield  {author} {\bibinfo {author} {\bibfnamefont {M.}~\bibnamefont
  {Milgrom}},\ }\href {\doibase 10.1103/PhysRevD.80.123536} {\bibfield
  {journal} {\bibinfo  {journal} {Phys. Rev. D}\ }\textbf {\bibinfo {volume}
  {80}},\ \bibinfo {pages} {123536} (\bibinfo {year} {2009})},\ \Eprint
  {http://arxiv.org/abs/0912.0790} {arXiv:0912.0790 [gr-qc]} \BibitemShut
  {NoStop}%
\bibitem [{\citenamefont {Skordis}\ and\ \citenamefont
  {Zlosnik}(2021{\natexlab{a}})}]{2007.00082}%
  \BibitemOpen
  \bibfield  {author} {\bibinfo {author} {\bibfnamefont {C.}~\bibnamefont
  {Skordis}}\ and\ \bibinfo {author} {\bibfnamefont {T.}~\bibnamefont
  {Zlosnik}},\ }\href {\doibase 10.1103/PhysRevLett.127.161302} {\bibfield
  {journal} {\bibinfo  {journal} {Phys. Rev. Lett.}\ }\textbf {\bibinfo
  {volume} {127}},\ \bibinfo {pages} {161302} (\bibinfo {year}
  {2021}{\natexlab{a}})},\ \Eprint {http://arxiv.org/abs/2007.00082}
  {arXiv:2007.00082 [astro-ph.CO]} \BibitemShut {NoStop}%
\bibitem [{\citenamefont {Skordis}\ and\ \citenamefont
  {Zlosnik}(2021{\natexlab{b}})}]{2109.13287}%
  \BibitemOpen
  \bibfield  {author} {\bibinfo {author} {\bibfnamefont {C.}~\bibnamefont
  {Skordis}}\ and\ \bibinfo {author} {\bibfnamefont {T.}~\bibnamefont
  {Zlosnik}},\ }\href@noop {} {\  (\bibinfo {year} {2021}{\natexlab{b}})},\
  \Eprint {http://arxiv.org/abs/2109.13287} {arXiv:2109.13287 [gr-qc]}
  \BibitemShut {NoStop}%
\bibitem [{\citenamefont {Nusser}(2002)}]{0109016}%
  \BibitemOpen
  \bibfield  {author} {\bibinfo {author} {\bibfnamefont {A.}~\bibnamefont
  {Nusser}},\ }\href {\doibase 10.1046/j.1365-8711.2002.05235.x} {\bibfield
  {journal} {\bibinfo  {journal} {Mon. Not. Roy. Astron. Soc.}\ }\textbf
  {\bibinfo {volume} {331}},\ \bibinfo {pages} {909} (\bibinfo {year}
  {2002})},\ \Eprint {http://arxiv.org/abs/astro-ph/0109016}
  {arXiv:astro-ph/0109016} \BibitemShut {NoStop}%
\bibitem [{\citenamefont {Knebe}\ and\ \citenamefont {Gibson}(2004)}]{0303222}%
  \BibitemOpen
  \bibfield  {author} {\bibinfo {author} {\bibfnamefont {A.}~\bibnamefont
  {Knebe}}\ and\ \bibinfo {author} {\bibfnamefont {B.~K.}\ \bibnamefont
  {Gibson}},\ }\href {\doibase 10.1111/j.1365-2966.2004.07182.x} {\bibfield
  {journal} {\bibinfo  {journal} {Mon. Not. Roy. Astron. Soc.}\ }\textbf
  {\bibinfo {volume} {347}},\ \bibinfo {pages} {1055} (\bibinfo {year}
  {2004})},\ \Eprint {http://arxiv.org/abs/astro-ph/0303222}
  {arXiv:astro-ph/0303222} \BibitemShut {NoStop}%
\bibitem [{\citenamefont {Llinares}\ \emph {et~al.}(2008)\citenamefont
  {Llinares}, \citenamefont {Knebe},\ and\ \citenamefont {Zhao}}]{0809.2899}%
  \BibitemOpen
  \bibfield  {author} {\bibinfo {author} {\bibfnamefont {C.}~\bibnamefont
  {Llinares}}, \bibinfo {author} {\bibfnamefont {A.}~\bibnamefont {Knebe}}, \
  and\ \bibinfo {author} {\bibfnamefont {H.}~\bibnamefont {Zhao}},\ }\href
  {\doibase 10.1111/j.1365-2966.2008.13961.x} {\bibfield  {journal} {\bibinfo
  {journal} {Mon. Not. Roy. Astron. Soc.}\ }\textbf {\bibinfo {volume} {391}},\
  \bibinfo {pages} {1778} (\bibinfo {year} {2008})},\ \Eprint
  {http://arxiv.org/abs/0809.2899} {arXiv:0809.2899 [astro-ph]} \BibitemShut
  {NoStop}%
\bibitem [{\citenamefont {{Angus}}\ and\ \citenamefont
  {{Diaferio}}(2011)}]{1104.5040}%
  \BibitemOpen
  \bibfield  {author} {\bibinfo {author} {\bibfnamefont {G.~W.}\ \bibnamefont
  {{Angus}}}\ and\ \bibinfo {author} {\bibfnamefont {A.}~\bibnamefont
  {{Diaferio}}},\ }\href {\doibase 10.1111/j.1365-2966.2011.19321.x} {\bibfield
   {journal} {\bibinfo  {journal} {\mnras}\ }\textbf {\bibinfo {volume}
  {417}},\ \bibinfo {pages} {941} (\bibinfo {year} {2011})},\ \Eprint
  {http://arxiv.org/abs/1104.5040} {arXiv:1104.5040 [astro-ph.CO]} \BibitemShut
  {NoStop}%
\bibitem [{\citenamefont {Angus}\ \emph {et~al.}(2013)\citenamefont {Angus},
  \citenamefont {Diaferio}, \citenamefont {Famaey},\ and\ \citenamefont
  {van~der Heyden}}]{1309.6094}%
  \BibitemOpen
  \bibfield  {author} {\bibinfo {author} {\bibfnamefont {G.~W.}\ \bibnamefont
  {Angus}}, \bibinfo {author} {\bibfnamefont {A.}~\bibnamefont {Diaferio}},
  \bibinfo {author} {\bibfnamefont {B.}~\bibnamefont {Famaey}}, \ and\ \bibinfo
  {author} {\bibfnamefont {K.~J.}\ \bibnamefont {van~der Heyden}},\ }\href
  {\doibase 10.1093/mnras/stt1564} {\bibfield  {journal} {\bibinfo  {journal}
  {Mon. Not. Roy. Astron. Soc.}\ }\textbf {\bibinfo {volume} {436}},\ \bibinfo
  {pages} {202} (\bibinfo {year} {2013})},\ \Eprint
  {http://arxiv.org/abs/1309.6094} {arXiv:1309.6094 [astro-ph.CO]} \BibitemShut
  {NoStop}%
\bibitem [{\citenamefont {Candlish}\ \emph {et~al.}(2015)\citenamefont
  {Candlish}, \citenamefont {Smith},\ and\ \citenamefont
  {Fellhauer}}]{1410.3844}%
  \BibitemOpen
  \bibfield  {author} {\bibinfo {author} {\bibfnamefont {G.~N.}\ \bibnamefont
  {Candlish}}, \bibinfo {author} {\bibfnamefont {R.}~\bibnamefont {Smith}}, \
  and\ \bibinfo {author} {\bibfnamefont {M.}~\bibnamefont {Fellhauer}},\ }\href
  {\doibase 10.1093/mnras/stu2158} {\bibfield  {journal} {\bibinfo  {journal}
  {Mon. Not. Roy. Astron. Soc.}\ }\textbf {\bibinfo {volume} {446}},\ \bibinfo
  {pages} {1060} (\bibinfo {year} {2015})},\ \Eprint
  {http://arxiv.org/abs/1410.3844} {arXiv:1410.3844 [astro-ph.GA]} \BibitemShut
  {NoStop}%
\bibitem [{\citenamefont {Candlish}(2016)}]{1605.03192}%
  \BibitemOpen
  \bibfield  {author} {\bibinfo {author} {\bibfnamefont {G.~N.}\ \bibnamefont
  {Candlish}},\ }\href {\doibase 10.1093/mnras/stw1130} {\bibfield  {journal}
  {\bibinfo  {journal} {Mon. Not. Roy. Astron. Soc.}\ }\textbf {\bibinfo
  {volume} {460}},\ \bibinfo {pages} {2571} (\bibinfo {year} {2016})},\ \Eprint
  {http://arxiv.org/abs/1605.03192} {arXiv:1605.03192 [astro-ph.CO]}
  \BibitemShut {NoStop}%
\bibitem [{\citenamefont {Milillo}\ \emph {et~al.}(2015)\citenamefont
  {Milillo}, \citenamefont {Bertacca}, \citenamefont {Bruni},\ and\
  \citenamefont {Maselli}}]{postf}%
  \BibitemOpen
  \bibfield  {author} {\bibinfo {author} {\bibfnamefont {I.}~\bibnamefont
  {Milillo}}, \bibinfo {author} {\bibfnamefont {D.}~\bibnamefont {Bertacca}},
  \bibinfo {author} {\bibfnamefont {M.}~\bibnamefont {Bruni}}, \ and\ \bibinfo
  {author} {\bibfnamefont {A.}~\bibnamefont {Maselli}},\ }\href {\doibase
  10.1103/PhysRevD.92.023519} {\bibfield  {journal} {\bibinfo  {journal} {Phys.
  Rev. D}\ }\textbf {\bibinfo {volume} {92}},\ \bibinfo {pages} {023519}
  (\bibinfo {year} {2015})},\ \Eprint {http://arxiv.org/abs/1502.02985}
  {arXiv:1502.02985 [gr-qc]} \BibitemShut {NoStop}%
\bibitem [{\citenamefont {Milillo}(2010)}]{thesis}%
  \BibitemOpen
  \bibfield  {author} {\bibinfo {author} {\bibfnamefont {I.}~\bibnamefont
  {Milillo}},\ }\emph {\bibinfo {title} {Linear and non-linear effects in
  structure formation}},\ \href@noop {} {Ph.D. thesis},\ \bibinfo  {school}
  {University of Portsmouth} (\bibinfo {year} {2010})\BibitemShut {NoStop}%
\bibitem [{\citenamefont {Hu}\ and\ \citenamefont {Sawicki}(2007)}]{husawicki}%
  \BibitemOpen
  \bibfield  {author} {\bibinfo {author} {\bibfnamefont {W.}~\bibnamefont
  {Hu}}\ and\ \bibinfo {author} {\bibfnamefont {I.}~\bibnamefont {Sawicki}},\
  }\href {\doibase 10.1103/PhysRevD.76.064004} {\bibfield  {journal} {\bibinfo
  {journal} {Phys. Rev.}\ }\textbf {\bibinfo {volume} {D76}},\ \bibinfo {pages}
  {064004} (\bibinfo {year} {2007})},\ \Eprint {http://arxiv.org/abs/0705.1158}
  {arXiv:0705.1158 [astro-ph]} \BibitemShut {NoStop}%
\bibitem [{\citenamefont {{Thomas}}\ \emph {et~al.}(2015)\citenamefont
  {{Thomas}}, \citenamefont {{Bruni}}, \citenamefont {{Koyama}}, \citenamefont
  {{Li}},\ and\ \citenamefont {{Zhao}}}]{postffr}%
  \BibitemOpen
  \bibfield  {author} {\bibinfo {author} {\bibfnamefont {D.~B.}\ \bibnamefont
  {{Thomas}}}, \bibinfo {author} {\bibfnamefont {M.}~\bibnamefont {{Bruni}}},
  \bibinfo {author} {\bibfnamefont {K.}~\bibnamefont {{Koyama}}}, \bibinfo
  {author} {\bibfnamefont {B.}~\bibnamefont {{Li}}}, \ and\ \bibinfo {author}
  {\bibfnamefont {G.-B.}\ \bibnamefont {{Zhao}}},\ }\href {\doibase
  10.1088/1475-7516/2015/07/051} {\bibfield  {journal} {\bibinfo  {journal}
  {\jcap}\ }\textbf {\bibinfo {volume} {7}},\ \bibinfo {eid} {051} (\bibinfo
  {year} {2015})},\ \Eprint {http://arxiv.org/abs/1503.07204} {arXiv:1503.07204
  [gr-qc]} \BibitemShut {NoStop}%
\bibitem [{\citenamefont {Thomas}(2020)}]{theorypaper}%
  \BibitemOpen
  \bibfield  {author} {\bibinfo {author} {\bibfnamefont {D.~B.}\ \bibnamefont
  {Thomas}},\ }\href {\doibase 10.1103/physrevd.101.123517} {\bibfield
  {journal} {\bibinfo  {journal} {Physical Review D}\ }\textbf {\bibinfo
  {volume} {101}} (\bibinfo {year} {2020}),\
  10.1103/physrevd.101.123517}\BibitemShut {NoStop}%
\bibitem [{\citenamefont {Srinivasan}\ \emph {et~al.}(2021)\citenamefont
  {Srinivasan}, \citenamefont {Thomas}, \citenamefont {Pace},\ and\
  \citenamefont {Battye}}]{simpaper}%
  \BibitemOpen
  \bibfield  {author} {\bibinfo {author} {\bibfnamefont {S.}~\bibnamefont
  {Srinivasan}}, \bibinfo {author} {\bibfnamefont {D.~B.}\ \bibnamefont
  {Thomas}}, \bibinfo {author} {\bibfnamefont {F.}~\bibnamefont {Pace}}, \ and\
  \bibinfo {author} {\bibfnamefont {R.}~\bibnamefont {Battye}},\ }\href
  {\doibase 10.1088/1475-7516/2021/06/016} {\bibfield  {journal} {\bibinfo
  {journal} {JCAP}\ }\textbf {\bibinfo {volume} {06}},\ \bibinfo {pages} {016}
  (\bibinfo {year} {2021})},\ \Eprint {http://arxiv.org/abs/2103.05051}
  {arXiv:2103.05051 [astro-ph.CO]} \BibitemShut {NoStop}%
\bibitem [{\citenamefont {Zlosnik}\ \emph {et~al.}(2007)\citenamefont
  {Zlosnik}, \citenamefont {Ferreira},\ and\ \citenamefont
  {Starkman}}]{0607411}%
  \BibitemOpen
  \bibfield  {author} {\bibinfo {author} {\bibfnamefont {T.~G.}\ \bibnamefont
  {Zlosnik}}, \bibinfo {author} {\bibfnamefont {P.~G.}\ \bibnamefont
  {Ferreira}}, \ and\ \bibinfo {author} {\bibfnamefont {G.~D.}\ \bibnamefont
  {Starkman}},\ }\href {\doibase 10.1103/PhysRevD.75.044017} {\bibfield
  {journal} {\bibinfo  {journal} {Phys. Rev.}\ }\textbf {\bibinfo {volume}
  {D75}},\ \bibinfo {pages} {044017} (\bibinfo {year} {2007})},\ \Eprint
  {http://arxiv.org/abs/astro-ph/0607411} {arXiv:astro-ph/0607411 [astro-ph]}
  \BibitemShut {NoStop}%
\bibitem [{\citenamefont {Dai}\ \emph {et~al.}(2008)\citenamefont {Dai},
  \citenamefont {Matsuo},\ and\ \citenamefont {Starkman}}]{Dai:2008sf}%
  \BibitemOpen
  \bibfield  {author} {\bibinfo {author} {\bibfnamefont {D.-C.}\ \bibnamefont
  {Dai}}, \bibinfo {author} {\bibfnamefont {R.}~\bibnamefont {Matsuo}}, \ and\
  \bibinfo {author} {\bibfnamefont {G.}~\bibnamefont {Starkman}},\ }\href
  {\doibase 10.1103/PhysRevD.78.104004} {\bibfield  {journal} {\bibinfo
  {journal} {Phys. Rev. D}\ }\textbf {\bibinfo {volume} {78}},\ \bibinfo
  {pages} {104004} (\bibinfo {year} {2008})},\ \Eprint
  {http://arxiv.org/abs/0806.4319} {arXiv:0806.4319 [gr-qc]} \BibitemShut
  {NoStop}%
\bibitem [{\citenamefont {Zlosnik}\ \emph {et~al.}(2008)\citenamefont
  {Zlosnik}, \citenamefont {Ferreira},\ and\ \citenamefont
  {Starkman}}]{0711.0520}%
  \BibitemOpen
  \bibfield  {author} {\bibinfo {author} {\bibfnamefont {T.~G.}\ \bibnamefont
  {Zlosnik}}, \bibinfo {author} {\bibfnamefont {P.~G.}\ \bibnamefont
  {Ferreira}}, \ and\ \bibinfo {author} {\bibfnamefont {G.~D.}\ \bibnamefont
  {Starkman}},\ }\href {\doibase 10.1103/PhysRevD.77.084010} {\bibfield
  {journal} {\bibinfo  {journal} {Phys. Rev. D}\ }\textbf {\bibinfo {volume}
  {77}},\ \bibinfo {pages} {084010} (\bibinfo {year} {2008})},\ \Eprint
  {http://arxiv.org/abs/0711.0520} {arXiv:0711.0520 [astro-ph]} \BibitemShut
  {NoStop}%
\bibitem [{\citenamefont {Zuntz}\ \emph {et~al.}(2010)\citenamefont {Zuntz},
  \citenamefont {Zlosnik}, \citenamefont {Bourliot}, \citenamefont {Ferreira},\
  and\ \citenamefont {Starkman}}]{1002.0849}%
  \BibitemOpen
  \bibfield  {author} {\bibinfo {author} {\bibfnamefont {J.}~\bibnamefont
  {Zuntz}}, \bibinfo {author} {\bibfnamefont {T.~G.}\ \bibnamefont {Zlosnik}},
  \bibinfo {author} {\bibfnamefont {F.}~\bibnamefont {Bourliot}}, \bibinfo
  {author} {\bibfnamefont {P.~G.}\ \bibnamefont {Ferreira}}, \ and\ \bibinfo
  {author} {\bibfnamefont {G.~D.}\ \bibnamefont {Starkman}},\ }\href {\doibase
  10.1103/PhysRevD.81.104015} {\bibfield  {journal} {\bibinfo  {journal} {Phys.
  Rev. D}\ }\textbf {\bibinfo {volume} {81}},\ \bibinfo {pages} {104015}
  (\bibinfo {year} {2010})},\ \Eprint {http://arxiv.org/abs/1002.0849}
  {arXiv:1002.0849 [astro-ph.CO]} \BibitemShut {NoStop}%
\bibitem [{\citenamefont {Li}\ \emph {et~al.}(2008)\citenamefont {Li},
  \citenamefont {Fonseca~Mota},\ and\ \citenamefont {Barrow}}]{0709.4581}%
  \BibitemOpen
  \bibfield  {author} {\bibinfo {author} {\bibfnamefont {B.}~\bibnamefont
  {Li}}, \bibinfo {author} {\bibfnamefont {D.}~\bibnamefont {Fonseca~Mota}}, \
  and\ \bibinfo {author} {\bibfnamefont {J.~D.}\ \bibnamefont {Barrow}},\
  }\href {\doibase 10.1103/PhysRevD.77.024032} {\bibfield  {journal} {\bibinfo
  {journal} {Phys. Rev. D}\ }\textbf {\bibinfo {volume} {77}},\ \bibinfo
  {pages} {024032} (\bibinfo {year} {2008})},\ \Eprint
  {http://arxiv.org/abs/0709.4581} {arXiv:0709.4581 [astro-ph]} \BibitemShut
  {NoStop}%
\bibitem [{\citenamefont {Lagos}\ \emph {et~al.}(2018)\citenamefont {Lagos},
  \citenamefont {Bellini}, \citenamefont {Noller}, \citenamefont {Ferreira},\
  and\ \citenamefont {Baker}}]{1711.09893}%
  \BibitemOpen
  \bibfield  {author} {\bibinfo {author} {\bibfnamefont {M.}~\bibnamefont
  {Lagos}}, \bibinfo {author} {\bibfnamefont {E.}~\bibnamefont {Bellini}},
  \bibinfo {author} {\bibfnamefont {J.}~\bibnamefont {Noller}}, \bibinfo
  {author} {\bibfnamefont {P.~G.}\ \bibnamefont {Ferreira}}, \ and\ \bibinfo
  {author} {\bibfnamefont {T.}~\bibnamefont {Baker}},\ }\href {\doibase
  10.1088/1475-7516/2018/03/021} {\bibfield  {journal} {\bibinfo  {journal}
  {JCAP}\ }\textbf {\bibinfo {volume} {1803}},\ \bibinfo {pages} {021}
  (\bibinfo {year} {2018})},\ \Eprint {http://arxiv.org/abs/1711.09893}
  {arXiv:1711.09893 [gr-qc]} \BibitemShut {NoStop}%
\bibitem [{\citenamefont {Battye}\ \emph {et~al.}(2017)\citenamefont {Battye},
  \citenamefont {Pace},\ and\ \citenamefont {Trinh}}]{1707.06508}%
  \BibitemOpen
  \bibfield  {author} {\bibinfo {author} {\bibfnamefont {R.~A.}\ \bibnamefont
  {Battye}}, \bibinfo {author} {\bibfnamefont {F.}~\bibnamefont {Pace}}, \ and\
  \bibinfo {author} {\bibfnamefont {D.}~\bibnamefont {Trinh}},\ }\href
  {\doibase 10.1103/PhysRevD.96.064041} {\bibfield  {journal} {\bibinfo
  {journal} {Phys. Rev. D}\ }\textbf {\bibinfo {volume} {96}},\ \bibinfo
  {pages} {064041} (\bibinfo {year} {2017})},\ \Eprint
  {http://arxiv.org/abs/1707.06508} {arXiv:1707.06508 [astro-ph.CO]}
  \BibitemShut {NoStop}%
\bibitem [{\citenamefont {Trinh}\ \emph {et~al.}(2019)\citenamefont {Trinh},
  \citenamefont {Pace}, \citenamefont {Battye},\ and\ \citenamefont
  {Bolliet}}]{1811.07805}%
  \BibitemOpen
  \bibfield  {author} {\bibinfo {author} {\bibfnamefont {D.}~\bibnamefont
  {Trinh}}, \bibinfo {author} {\bibfnamefont {F.}~\bibnamefont {Pace}},
  \bibinfo {author} {\bibfnamefont {R.~A.}\ \bibnamefont {Battye}}, \ and\
  \bibinfo {author} {\bibfnamefont {B.}~\bibnamefont {Bolliet}},\ }\href
  {\doibase 10.1103/PhysRevD.99.043515} {\bibfield  {journal} {\bibinfo
  {journal} {Phys. Rev. D}\ }\textbf {\bibinfo {volume} {99}},\ \bibinfo
  {pages} {043515} (\bibinfo {year} {2019})},\ \Eprint
  {http://arxiv.org/abs/1811.07805} {arXiv:1811.07805 [astro-ph.CO]}
  \BibitemShut {NoStop}%
\bibitem [{\citenamefont {Sanghai}\ and\ \citenamefont
  {Clifton}(2017)}]{ppnc1}%
  \BibitemOpen
  \bibfield  {author} {\bibinfo {author} {\bibfnamefont {V.~A.~A.}\
  \bibnamefont {Sanghai}}\ and\ \bibinfo {author} {\bibfnamefont
  {T.}~\bibnamefont {Clifton}},\ }\href {\doibase 10.1088/1361-6382/aa5d75}
  {\bibfield  {journal} {\bibinfo  {journal} {Classical and Quantum Gravity}\
  }\textbf {\bibinfo {volume} {34}},\ \bibinfo {pages} {065003} (\bibinfo
  {year} {2017})}\BibitemShut {NoStop}%
\bibitem [{\citenamefont {Clifton}\ and\ \citenamefont
  {Sanghai}(2019)}]{ppnc2}%
  \BibitemOpen
  \bibfield  {author} {\bibinfo {author} {\bibfnamefont {T.}~\bibnamefont
  {Clifton}}\ and\ \bibinfo {author} {\bibfnamefont {V.~A.~A.}\ \bibnamefont
  {Sanghai}},\ }\href {\doibase 10.1103/PhysRevLett.122.011301} {\bibfield
  {journal} {\bibinfo  {journal} {Phys. Rev. Lett.}\ }\textbf {\bibinfo
  {volume} {122}},\ \bibinfo {pages} {011301} (\bibinfo {year}
  {2019})}\BibitemShut {NoStop}%
\bibitem [{\citenamefont {Sanders}(1998)}]{9710335}%
  \BibitemOpen
  \bibfield  {author} {\bibinfo {author} {\bibfnamefont {R.~H.}\ \bibnamefont
  {Sanders}},\ }\href {\doibase 10.1046/j.1365-8711.1998.01459.x} {\bibfield
  {journal} {\bibinfo  {journal} {Mon. Not. Roy. Astron. Soc.}\ }\textbf
  {\bibinfo {volume} {296}},\ \bibinfo {pages} {1009} (\bibinfo {year}
  {1998})},\ \Eprint {http://arxiv.org/abs/astro-ph/9710335}
  {arXiv:astro-ph/9710335} \BibitemShut {NoStop}%
\bibitem [{\citenamefont {Sanders}(2001)}]{0011439}%
  \BibitemOpen
  \bibfield  {author} {\bibinfo {author} {\bibfnamefont {R.~H.}\ \bibnamefont
  {Sanders}},\ }\href {\doibase 10.1086/322487} {\bibfield  {journal} {\bibinfo
   {journal} {Astrophys. J.}\ }\textbf {\bibinfo {volume} {560}},\ \bibinfo
  {pages} {1} (\bibinfo {year} {2001})},\ \Eprint
  {http://arxiv.org/abs/astro-ph/0011439} {arXiv:astro-ph/0011439} \BibitemShut
  {NoStop}%
\bibitem [{\citenamefont {Baker}\ \emph {et~al.}(2017)\citenamefont {Baker},
  \citenamefont {Bellini}, \citenamefont {Ferreira}, \citenamefont {Lagos},
  \citenamefont {Noller},\ and\ \citenamefont {Sawicki}}]{1710.06394}%
  \BibitemOpen
  \bibfield  {author} {\bibinfo {author} {\bibfnamefont {T.}~\bibnamefont
  {Baker}}, \bibinfo {author} {\bibfnamefont {E.}~\bibnamefont {Bellini}},
  \bibinfo {author} {\bibfnamefont {P.~G.}\ \bibnamefont {Ferreira}}, \bibinfo
  {author} {\bibfnamefont {M.}~\bibnamefont {Lagos}}, \bibinfo {author}
  {\bibfnamefont {J.}~\bibnamefont {Noller}}, \ and\ \bibinfo {author}
  {\bibfnamefont {I.}~\bibnamefont {Sawicki}},\ }\href {\doibase
  10.1103/PhysRevLett.119.251301} {\bibfield  {journal} {\bibinfo  {journal}
  {Phys. Rev. Lett.}\ }\textbf {\bibinfo {volume} {119}},\ \bibinfo {pages}
  {251301} (\bibinfo {year} {2017})},\ \Eprint
  {http://arxiv.org/abs/1710.06394} {arXiv:1710.06394 [astro-ph.CO]}
  \BibitemShut {NoStop}%
\bibitem [{\citenamefont {Battye}\ \emph {et~al.}(2018)\citenamefont {Battye},
  \citenamefont {Pace},\ and\ \citenamefont {Trinh}}]{1802.09447}%
  \BibitemOpen
  \bibfield  {author} {\bibinfo {author} {\bibfnamefont {R.~A.}\ \bibnamefont
  {Battye}}, \bibinfo {author} {\bibfnamefont {F.}~\bibnamefont {Pace}}, \ and\
  \bibinfo {author} {\bibfnamefont {D.}~\bibnamefont {Trinh}},\ }\href
  {\doibase 10.1103/PhysRevD.98.023504} {\bibfield  {journal} {\bibinfo
  {journal} {Phys. Rev. D}\ }\textbf {\bibinfo {volume} {98}},\ \bibinfo
  {pages} {023504} (\bibinfo {year} {2018})},\ \Eprint
  {http://arxiv.org/abs/1802.09447} {arXiv:1802.09447 [astro-ph.CO]}
  \BibitemShut {NoStop}%
\bibitem [{\citenamefont {Clifton}\ \emph {et~al.}(2020)\citenamefont
  {Clifton}, \citenamefont {Gallagher}, \citenamefont {Goldberg},\ and\
  \citenamefont {Malik}}]{cliftongauge}%
  \BibitemOpen
  \bibfield  {author} {\bibinfo {author} {\bibfnamefont {T.}~\bibnamefont
  {Clifton}}, \bibinfo {author} {\bibfnamefont {C.~S.}\ \bibnamefont
  {Gallagher}}, \bibinfo {author} {\bibfnamefont {S.}~\bibnamefont {Goldberg}},
  \ and\ \bibinfo {author} {\bibfnamefont {K.~A.}\ \bibnamefont {Malik}},\
  }\href {\doibase 10.1103/PhysRevD.101.063530} {\bibfield  {journal} {\bibinfo
   {journal} {Phys. Rev. D}\ }\textbf {\bibinfo {volume} {101}},\ \bibinfo
  {pages} {063530} (\bibinfo {year} {2020})},\ \Eprint
  {http://arxiv.org/abs/2001.00394} {arXiv:2001.00394 [gr-qc]} \BibitemShut
  {NoStop}%
\bibitem [{\citenamefont {Bruni}\ \emph {et~al.}(2014)\citenamefont {Bruni},
  \citenamefont {Thomas},\ and\ \citenamefont {Wands}}]{Bruni:2013mua}%
  \BibitemOpen
  \bibfield  {author} {\bibinfo {author} {\bibfnamefont {M.}~\bibnamefont
  {Bruni}}, \bibinfo {author} {\bibfnamefont {D.~B.}\ \bibnamefont {Thomas}}, \
  and\ \bibinfo {author} {\bibfnamefont {D.}~\bibnamefont {Wands}},\ }\href
  {\doibase 10.1103/PhysRevD.89.044010} {\bibfield  {journal} {\bibinfo
  {journal} {Phys.Rev.}\ }\textbf {\bibinfo {volume} {D89}},\ \bibinfo {pages}
  {044010} (\bibinfo {year} {2014})},\ \Eprint {http://arxiv.org/abs/1306.1562}
  {arXiv:1306.1562} \BibitemShut {NoStop}%
\bibitem [{\citenamefont {Thomas}\ \emph {et~al.}(2015)\citenamefont {Thomas},
  \citenamefont {Bruni},\ and\ \citenamefont {Wands}}]{longerpaper}%
  \BibitemOpen
  \bibfield  {author} {\bibinfo {author} {\bibfnamefont {D.~B.}\ \bibnamefont
  {Thomas}}, \bibinfo {author} {\bibfnamefont {M.}~\bibnamefont {Bruni}}, \
  and\ \bibinfo {author} {\bibfnamefont {D.}~\bibnamefont {Wands}},\ }\href
  {\doibase 10.1093/mnras/stv1390} {\bibfield  {journal} {\bibinfo  {journal}
  {Mon. Not. Roy. Astron. Soc.}\ }\textbf {\bibinfo {volume} {452}},\ \bibinfo
  {pages} {1727} (\bibinfo {year} {2015})},\ \Eprint
  {http://arxiv.org/abs/1501.00799} {arXiv:1501.00799 [astro-ph.CO]}
  \BibitemShut {NoStop}%
\bibitem [{\citenamefont {Avilez-Lopez}\ \emph {et~al.}(2015)\citenamefont
  {Avilez-Lopez}, \citenamefont {Padilla}, \citenamefont {Saffin},\ and\
  \citenamefont {Skordis}}]{PPNV}%
  \BibitemOpen
  \bibfield  {author} {\bibinfo {author} {\bibfnamefont {A.}~\bibnamefont
  {Avilez-Lopez}}, \bibinfo {author} {\bibfnamefont {A.}~\bibnamefont
  {Padilla}}, \bibinfo {author} {\bibfnamefont {P.~M.}\ \bibnamefont {Saffin}},
  \ and\ \bibinfo {author} {\bibfnamefont {C.}~\bibnamefont {Skordis}},\ }\href
  {\doibase 10.1088/1475-7516/2015/06/044} {\bibfield  {journal} {\bibinfo
  {journal} {JCAP}\ }\textbf {\bibinfo {volume} {06}},\ \bibinfo {pages} {044}
  (\bibinfo {year} {2015})},\ \Eprint {http://arxiv.org/abs/1501.01985}
  {arXiv:1501.01985 [gr-qc]} \BibitemShut {NoStop}%
\bibitem [{\citenamefont {{Meng}}\ and\ \citenamefont
  {{Du}}(2012)}]{gealogfunc}%
  \BibitemOpen
  \bibfield  {author} {\bibinfo {author} {\bibfnamefont {X.}~\bibnamefont
  {{Meng}}}\ and\ \bibinfo {author} {\bibfnamefont {X.}~\bibnamefont {{Du}}},\
  }\href {\doibase 10.1016/j.physletb.2012.03.024} {\bibfield  {journal}
  {\bibinfo  {journal} {Physics Letters B}\ }\textbf {\bibinfo {volume}
  {710}},\ \bibinfo {pages} {493} (\bibinfo {year} {2012})}\BibitemShut
  {NoStop}%
\bibitem [{\citenamefont {Bekenstein}\ and\ \citenamefont
  {Sagi}(2008)}]{0802.1526}%
  \BibitemOpen
  \bibfield  {author} {\bibinfo {author} {\bibfnamefont {J.~D.}\ \bibnamefont
  {Bekenstein}}\ and\ \bibinfo {author} {\bibfnamefont {E.}~\bibnamefont
  {Sagi}},\ }\href {\doibase 10.1103/PhysRevD.77.103512} {\bibfield  {journal}
  {\bibinfo  {journal} {Phys. Rev. D}\ }\textbf {\bibinfo {volume} {77}},\
  \bibinfo {pages} {103512} (\bibinfo {year} {2008})},\ \Eprint
  {http://arxiv.org/abs/0802.1526} {arXiv:0802.1526 [astro-ph]} \BibitemShut
  {NoStop}%
\bibitem [{\citenamefont {Angus}(2009)}]{0805.4014}%
  \BibitemOpen
  \bibfield  {author} {\bibinfo {author} {\bibfnamefont {G.~W.}\ \bibnamefont
  {Angus}},\ }\href {\doibase 10.1111/j.1365-2966.2008.14341.x} {\bibfield
  {journal} {\bibinfo  {journal} {Mon. Not. Roy. Astron. Soc.}\ }\textbf
  {\bibinfo {volume} {394}},\ \bibinfo {pages} {527} (\bibinfo {year}
  {2009})},\ \Eprint {http://arxiv.org/abs/0805.4014} {arXiv:0805.4014
  [astro-ph]} \BibitemShut {NoStop}%
\bibitem [{\citenamefont {Thomas}\ \emph {et~al.}(2017)\citenamefont {Thomas},
  \citenamefont {Whittaker}, \citenamefont {Camera},\ and\ \citenamefont
  {Brown}}]{rotationest}%
  \BibitemOpen
  \bibfield  {author} {\bibinfo {author} {\bibfnamefont {D.~B.}\ \bibnamefont
  {Thomas}}, \bibinfo {author} {\bibfnamefont {L.}~\bibnamefont {Whittaker}},
  \bibinfo {author} {\bibfnamefont {S.}~\bibnamefont {Camera}}, \ and\ \bibinfo
  {author} {\bibfnamefont {M.~L.}\ \bibnamefont {Brown}},\ }\href {\doibase
  10.1093/mnras/stx1468} {\bibfield  {journal} {\bibinfo  {journal} {Mon. Not.
  Roy. Astron. Soc.}\ }\textbf {\bibinfo {volume} {470}},\ \bibinfo {pages}
  {3131} (\bibinfo {year} {2017})},\ \Eprint {http://arxiv.org/abs/1612.01533}
  {arXiv:1612.01533 [astro-ph.CO]} \BibitemShut {NoStop}%
\bibitem [{\citenamefont {Kopp}\ \emph {et~al.}(2018)\citenamefont {Kopp},
  \citenamefont {Skordis}, \citenamefont {Thomas},\ and\ \citenamefont
  {Ili\'c}}]{gdmwtime}%
  \BibitemOpen
  \bibfield  {author} {\bibinfo {author} {\bibfnamefont {M.}~\bibnamefont
  {Kopp}}, \bibinfo {author} {\bibfnamefont {C.}~\bibnamefont {Skordis}},
  \bibinfo {author} {\bibfnamefont {D.~B.}\ \bibnamefont {Thomas}}, \ and\
  \bibinfo {author} {\bibfnamefont {S.}~\bibnamefont {Ili\'c}},\ }\href
  {\doibase 10.1103/PhysRevLett.120.221102} {\bibfield  {journal} {\bibinfo
  {journal} {Phys. Rev. Lett.}\ }\textbf {\bibinfo {volume} {120}},\ \bibinfo
  {pages} {221102} (\bibinfo {year} {2018})},\ \Eprint
  {http://arxiv.org/abs/1802.09541} {arXiv:1802.09541 [astro-ph.CO]}
  \BibitemShut {NoStop}%
\bibitem [{\citenamefont {Pardo}\ and\ \citenamefont
  {Spergel}(2020)}]{spergel}%
  \BibitemOpen
  \bibfield  {author} {\bibinfo {author} {\bibfnamefont {K.}~\bibnamefont
  {Pardo}}\ and\ \bibinfo {author} {\bibfnamefont {D.~N.}\ \bibnamefont
  {Spergel}},\ }\href {\doibase 10.1103/PhysRevLett.125.211101} {\bibfield
  {journal} {\bibinfo  {journal} {Phys. Rev. Lett.}\ }\textbf {\bibinfo
  {volume} {125}},\ \bibinfo {pages} {211101} (\bibinfo {year} {2020})},\
  \Eprint {http://arxiv.org/abs/2007.00555} {arXiv:2007.00555 [astro-ph.CO]}
  \BibitemShut {NoStop}%
\end{thebibliography}%
\bibliographystyle{apsrev4-1}
\end{document}